\documentclass{aa501}

\usepackage{epsf,epsfig}

\begin{document}

%\thesaurus{08      % A&A Section xx: stars
%          (02.01.2 %Accretion
%           08.02.1 %Stars: close binaries,
%           08.09.2 %Stars: individual: V\,709 Cas
%           13.25.5 %X-rays: stars
%          )}
%
\title{ The X-ray emission  of the  Intermediate Polar 
%RX\,J0028.8+5917}
V\,709 Cas}

%\thanks{Based
%on observations collected with the NASA/ESA Hubble Space Telescope,
%obtained
%at the Space Telescope Science Institute and with the International
%Ultraviolet
%Explorer, obtained from the IUE Final Archive at VILSPA}}

\author{ D. de Martino\inst{1},
G. Matt\inst{2},
K. Mukai\inst{3,4}
T. Belloni\inst{5},
%K. Beuermann\inst{4},
J.M. Bonnet-Bidaud\inst{6},
L. Chiappetti\inst{7},
B.T. G\"ansicke\inst{8},
F. Haberl\inst{9},
M. Mouchet\inst{10,11}
%J. Patterson\inst{12}
}

\offprints{D.~de Martino}

\institute{
Osservatorio Astronomico di Capodimonte, Via Moiariello 16, I-80131 Napoli, 
Italy
\email{demartino@na.astro.it}
   \and
Dipartimento di Fisica, Universita' degli Studi Roma Tre, Via della
Vasca Navale, ROMA, Italy
\email{matt@fis.uniroma3.it}
\and
Laboratory for High Energy Physics, NASA/GSFC, Code 662, Greenbelt, MD
20771,USA
\and
Universities Space Research Association 
\email{mukai@milkyway.gsfc.nasa.gov}
  \and
Osservatorio Astronomico di Brera, Via E. Bianchi 46, I-23807 Merate,
Italy
\email{belloni@merate.mi.astro.it}
   \and
Service  d'Astrophysique Saclay Gif-Sur-Yvette, France
\email{bobi@discovery.saclay.cea.fr}
   \and
Istituto di Fisica Cosmica del CNR, Via Bassini 15, Milano, Italy
\email{lucio@ifctr.mi.cnr.it}
\and
Universit\"ats Sternwarte G\"ottingen, Geismarlandstr. 11, D-37083,
G\"ottingen, Germany
\email{boris@uni-sw.gwdg.de}
 \and
 Max Planck Institut f\"ur Extraterrestrische Physik Garching, Germany
\email{fwh@xray.mpe.mpg.de}
\and
DAEC et UMR 8631 du CNRS, Observatoire de Paris, Section de Meudon,
F-92195 Meudon Cedex,
France
\and
Universit\'e Denis Diderot, Place Jussieu F-75005, France
\email{martine.mouchet@obspm.fr}
%\and
%Department of Astronomy, Columbia University, NY1002, USA
%\email{jop@astro.columbia.edu}
}

\date{Received April 23, 2001; Accepted July 24, 2001}

\authorrunning{de Martino et al.}
\titlerunning{The X-ray emission ...}
\markboth{...}{...}

\abstract{ 
We present RXTE and BeppoSAX observations of the Intermediate
Polar V\,709 Cas acquired in 1997 and 1998 respectively. The X-ray
emission from 
0.1 to 30\,keV is dominated by the strong pulsation at the rotational
period of the white dwarf (312.8\,s) with no sign of orbital or
sideband periodicity, thus confirming previous ROSAT results. 
However, we detect changes in the power spectra between
the two epochs. While the second harmonic of the spin
period is present during both observations, the first harmonic is
absent in 1997.   An increase in the amplitude of the spin
pulsation is found between 1997 and 1998 together with a decrease in
the X--ray flux. The average X--ray spectrum from 0.1 to 100\,keV is well
described
by an isothermal plasma at $\sim$ 27\,keV plus complex absorption and an
iron K$_{\alpha}$ fluorescent line, due to reflection from the
white dwarf surface. The rotational pulsation is compatible with complex
absorption dominating the low energy range, while the high energy 
spin modulation can be attributed  to tall shocks above the 
accreting poles. The RXTE spectrum in 1997 also shows 
the presence of an absorption edge from ionized iron likely 
located in the pre--shock accretion flow.  The variations along the
spin period of the partial covering absorber and of reflection are 
compatible with the classical accretion curtain scenario. The variations
in the spin pulse characteristics and X--ray flux indicate that
V\,709 Cas experiences changes in the mass accretion rate on timescales
from months to years.
}

\maketitle 

\section{Introduction}

Intermediate Polars (IPs) are a subclass of magnetic Cataclysmic 
Variables (mCVs) consisting of  a weakly magnetized  (up to a few MG) and
asynchronously rotating ($\rm P_{\rm rot} < P_{\rm orb}$) white dwarf
(WD) which accretes material from 
a late type, main sequence, Roche-lobe filling secondary star.

\noindent The magnetic field influences the details of the flow down to
the WD poles. In low field IPs, an accretion disc,
truncated at the  magnetospheric radius, may form and the material flows
from the inner edge of the disc along the field lines in an arc-shaped
curtain towards the polar regions of the WD (Rosen et al. 1988). A
disc-less (or stream-fed)  accretion is expected in systems with higher
fields, where material accretes directly from the stream flowing
from the secondary star. A mixture of stream-fed and disc
accretion are also possible, where matter from the stream overpasses 
the disc (disc-overflow, Hellier et al. 1995).
Different configurations (disc-fed and disc-overflow) can be present at
different epochs in the same system, due to long term changes in the mass
accretion rate (de Martino et al. 1995; Buckley et al. 1996; de
Martino et al. 1999). 

\noindent The inflow of accreting material onto the polar caps
produces  
a stand-off shock, below which material
cools via thermal Bremsstrahlung ($\rm kT \sim  5-30\,keV$).
Hard
X-rays are highly absorbed by cold matter (column densities up to
$\rm 10^{23}\,cm^{-2}$) 
within the accretion flow (Ishida et al. 1994) and are
expected to be reflected by the WD atmosphere (Beardmore et al. 1995; Done
et al. 1995; Matt 1999).

\noindent  Because of their asynchronous rotation, IPs show a wide range
of
periodicities
at the WD spin ($\omega$), the orbital
($\Omega$) and sideband frequencies (Warner 1986; Patterson 1994, Warner 
1995), whose amplitudes can be wavelength dependent (de Martino
1993).
\noindent The X-ray periodicities provide direct information on the actual
accretion mode (Wynn \& King 1992; Norton et al. 1996),  disc-fed systems
being dominated by the WD rotational period. The
additional presence of periodicities, at the beat (or synodic) $\omega -
\Omega$ and orbital
($\Omega$) frequencies are an indication of disk-overflow accretion. A 
pure stream-fed system should show pulses at the synodic
frequency only (Hellier 1999). To date only V\,2400 Oph
is a confirmed stream--fed IP (Buckley et al. 1997).
Furthermore disc-fed systems  show different rotational
pulse characteristics: double-peaked X-ray pulses, generally observed in 
fast rotators such as  YY\,Dra,  can originate from the emission 
of two accreting poles with tall shocks or large
footprints (Norton et al. 1999), whilst single-peaked curves,
observed in  most slow rotators such as FO\,Aqr, 
can  instead be produced by small accreting regions. 
It has been suggested (Norton et al. 1999) that such differences
are due to the effect of the magnetic field strength, the rapid rotators 
containing low field WDs (Norton et al. 1999). Unfortunately, for IPs
such a proposal is difficult to confim observationally, due to
the lack of detectable optical-IR polarization in most cases and of
spectral features of the WD. An outstanding exception is
V\,709 Cas (Bonnet-Bidaud et al. 2001), the first  
IP known to date to display the WD atmospheric features, from which 
a magnetic field strength lower than 3\,MG has been estimated.

V\,709 Cas (RX\,J0028.8+5917) was identified as an IP
from the ROSAT All Sky Survey (Haberl \& Motch 1995; Motch et al. 1996).
The X-ray emission was found to be hard 
(HR2=[1.0-2.4]\,keV -- [0.4-1.0]\,keV/[0.4 - 2.4]\,keV) = +0.21) and
pulsed at a period of 312.8\,s  in pointed follow-up 
PSPC and HRI ROSAT observations (Haberl \& Motch 1995; Norton et
al. 1999) (see Table\,1 for the history of pointed X--ray
observations).
 Searches in the  HEAO-1,
UHURU and Ariel\,V catalogues lead to the identification of
V\,709 Cas with the previously detected hard X-ray sources
1H\,0025+588, 4U\,0027+59 and 3A\,0026+593 (Motch et al. 1996). 
%In
%particular  from  Ariel\,V data, intensity variations (outbursts or
%flares) up  to a factor of 20 have been reported (Warwick et al. 1981). 
The X-ray timing analysis shows the presence of the dominant
312.8\,s pulsation and its first and second harmonics, with no sign
of orbital and sideband variabilities (Norton et al. 1999).
Two possible orbital periods  at $\rm P_{\Omega}$=5.4\,h and $\rm
P_{\Omega}$=4.5\,h have been  inferred from low temporal resolution  
optical spectroscopy (Motch et al. 1996), each being the one day alias of
the other. The former has been recently confirmed 
spectroscopically (Bonnet-Bidaud et al. 2001) and photometrically 
(Kozhevnikov  2001). This suggests that V\,709 Cas 
is a classical IP: a hard X-ray emitting and disc--accreting mCV. 

In this paper we present X--ray observations of
V\,709 Cas acquired with the RossiXTE (RXTE) and the BeppoSAX
satellites.
The latter also provides the first simultaneous soft and hard X--ray
data which allow a final characterization of its X--ray spectrum and 
its variability.

\begin{table*}[t]     %  Table  1
\centering 
\caption{History of pointed X--ray observations of V\,709 Cas.}
\vspace{0.05in}
\begin{tabular}{lccccc}
\hline
\hline
%\noalign{\smallskip}
   ~ &          ~ &           ~ &           ~ &    ~ & ~ \cr
Date & Instrument & Energy Band & Count Rate  & Flux & Source \cr
 &       &  (keV) &  cts\,s$^{-1}$ & erg\,cm$^{-2}\,s^{-1}$ & ~ \cr
   ~ &          ~ &           ~ &           ~ &    ~ & ~ \cr
\noalign {\hrule}
   ~ &          ~ &           ~ &           ~ &    ~ & ~ \cr
1992, Jul. & {\em ROSAT\,PSPC} & 0.1--2.4 &  0.60 & 8.8$\times 10^{-12}$
&
{\em a} \cr
1997, Mar. & {\em RXTE\,PCA} & 2--10 & 5.20 & 6.6$\times 10^{-11}$ & {\em 
b} \cr
1998, Feb. & {\em ROSAT\,HRI} & 0.1--2.4 & 0.26 & $\sim$ 1$\times
10^{-11}$ &
{\em c} \cr
1998, Jul. & {\em BeppoSAX\,MECS} & 2--10 & 0.50 & 4.5$\times 10^{-11}$ &
{\em b}\cr
   ~ &          ~ &           ~ &           ~ &    ~ & ~ \cr
\hline
\hline
\end{tabular}
~\par
\begin{flushleft}
{\em a}: Haberl \& Motch (1995).\par
{\em b}: This work. \par
{\em c}: Norton et al. (1999). {\em Note:} The X--ray flux is
estimated using the count rate conversion facility  {\it PIMMS-HEASARC}
at GSFC-NASA, adopting kT=10\,keV and $\rm N_{H}=9.8\times
10^{20}\,cm^{-2}$. 
\end{flushleft}
\end{table*}

\section{Observations and data reduction}

\subsection{The RXTE data}

V\,709 Cas was observed with RXTE (Bradt et al. 1993) between March
26--28, 1997,
over  a 2.3 day interval.  As the source was not
significantly
detected with the HEXTE instrument, operating in the energy range 
20-200\,keV, we limited the analysis to the PCA instrument,
consisting of
an array of 5 proportional counter units (PCU) sensitive in the
2.5--60.0\,keV band. The effective on--source exposure time, after 
standard screening, was 40\,ks.  V\,709 Cas was
detected at a  net count rate of 
7.04$\pm$0.01\,cts\,s$^{-1}$ per PCU in the 2.5--25\,keV band. 
We have estimated the PCA background using  a faint source model
appropriate for the epoch of the observation.
The RXTE data are telemetred in  two standard, and up to four
additional, data modes.
% providing flexibility to meet various
%scientific needs without necessarily requiring a huge telemetry  rate.  
Two separate observation modes were used for spectral and timing
analysis.  For the production of spectra, we used {\tt Standard\,2} mode
data, which have a time bin size of 16 seconds. In order to improve the
signal to noise, we used only data from the top layers of the PCUs. This
reduces the net signal by 20$\%$ and the estimated background by roughly
50$\%$. Since spectral uncertainties, particularly above 10\,keV,
are dominated by systematic uncertainties in the background model, 
this improves the quality of the net spectrum. For timing analysis, 
we have used the {\tt Event} mode data,  with a 8 ms time resolution and
32 PHA channels. As no selection by PCU or PCU layer is possible with this
data mode, all layers from all PCUs were used. All analyzed data refer
to the effective 2.9--24.2 keV band.  The average background level in the
five PCU units is 77.1\,cts\,s$^{-1}$ from all layers, while it is
40.3\,cts\,s$^{-1}$ from the top layer. 

%The effective energy range on
%which the analysis has been
%performed is 2.9--24.2\,keV. 

\subsection{The BeppoSAX data}
 
A later observation with the BeppoSAX satellite (Boella et al. 1997) was
carried out
between 1998 July 5-7  with the
set of the co-aligned Narrow Fields Instruments (NFI),
used as prime pointing instruments, co\-ve\-ring the wide energy 
range  0.1 - 300\,keV.
 The source was detected 
with the LECS (0.1-10\,keV), the MECS (1.3-10\,keV) and PDS
(15-300\,keV) detectors at net count rates of
0.243$\pm$0.003\,cts\,s$^{-1}$, 
0.497$\pm$0.002\,cts\,s$^{-1}$ and 0.346$\pm$0.028\,cts\,s$^{-1}$  
respectively. Effective
on--source exposure times have been 35.3 \,ks for the LECS,
86.9\,ks for the MECS and  79.5\,ks for the PDS detector (two units, 
see below).

\noindent The LECS light curves and spectra have 
been extracted using a circular region
with a  radius of 8\,arcmin centred on the source, whilst
MECS light curves and spectra have been extracted with a smaller 4\,arcmin
radius.
While the whole LECS band will be used for timing analysis, the
spectra from this instrument have been analyzed only below 4\,keV due to
calibration problems at higher energies.  For both intruments, the
background has been measured and subtracted using the same detector
regions during blank field pointings. The measured background in the
LECS (2.74$\rm \times 10^{-2}\,cts\,s^{-1}$) and MECS (6.95$\rm \times
10^{-3}\,cts\,s^{-1}$) is much lower than the source count rate.
 
\noindent The PDS is a collimated instrument which monitors the
background continuously switching two (of the four) detectors with a
dwell time of 96\,s.
The PDS light curves and spectra have been extracted using a standard
routine provided by the BeppoSAX Data Center (SDC). We have
conservatively extracted the light curves in the energy range of
15--70\,keV, above which the source is hardly detected.
The average net count rate  in this range is
0.270$\pm$ 0.020\, cts\,s$^{-1}$.
The procedure allows  rejection of particle background events,
as well as spikes caused by single particle hits, which produce
 fluorescent cascades which are mainly recorded below
30\,keV. The
background spectrum  is evaluated for each of
the two half arrays accumulating the off--source spectra. The 
background light curves are constructed by linear interpolation between
two off--source pointings in each array. Then the spectra and light curves
from the two arrays are merged together to construct the 
background spectrum and light curve for final subtraction.
The average background count rate is 10.31\,cts\,s$^{-1}$.
We also performed 
alternative ways to subtract background, accounting for collimator 
dwell time  effects  in order to construct the net PDS light
curve for periodicity search purposes.  We detected no
significant modulation in the 15--70\,keV energy range (see sect.\,3). 

 The history of the pointed X--ray observations of V\,709 Cas is
reported in Table\,1.

% A mean
%"running average" background in each of the four units as the
%source+background
%ligth curve has been derived by defining a
%time tolerance (two and four collimator cycles) such that, for each time
%bin in the source + background light curve, all bins in the background
%light curve, which are within plus or minus the tolerance are used to
%compute a mean count rate and error statistics. For each unit the average
%backgrounds are combined to form a mean  

\section{Timing analysis}

A search for periodicities has been performed on  
light curves extracted from both data sets in similar energy bands. The
range 2.9-9.8\,keV 
has been used for the RXTE PCA data, and the 1.3-10\,keV
band for the BeppoSAX MECS data. In Fig.\,1. the RXTE
and BeppoSAX light curves with a temporal resolution of 32\,s are shown.
A short term periodicity  is clearly apparent, with
no indication of variability (periodic  or not) on a longer
(hours) timescale.
 
\begin{figure*}
\begin{center}
\mbox{\epsfxsize=9cm\epsfbox{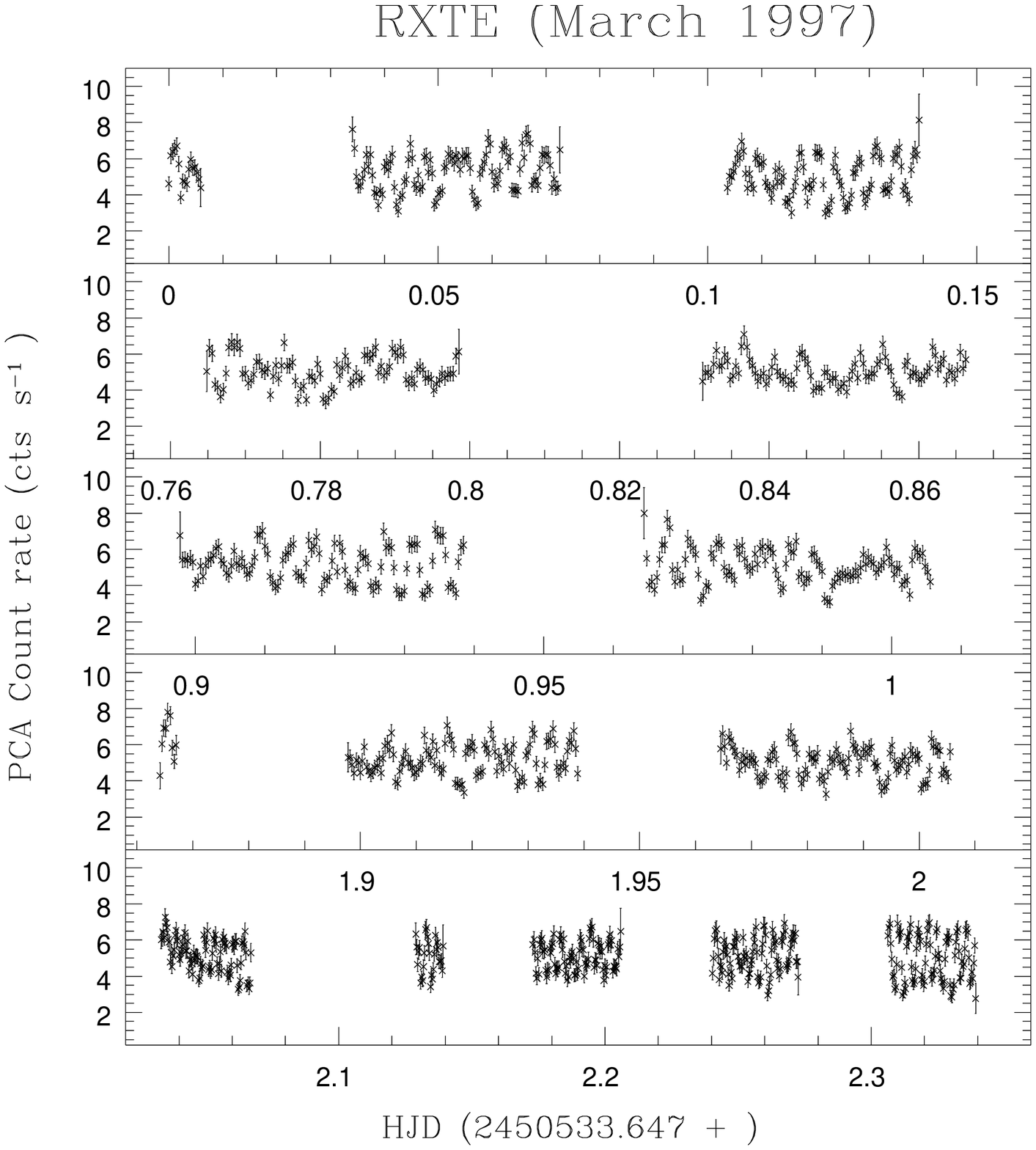}\epsfxsize=9cm\epsfbox{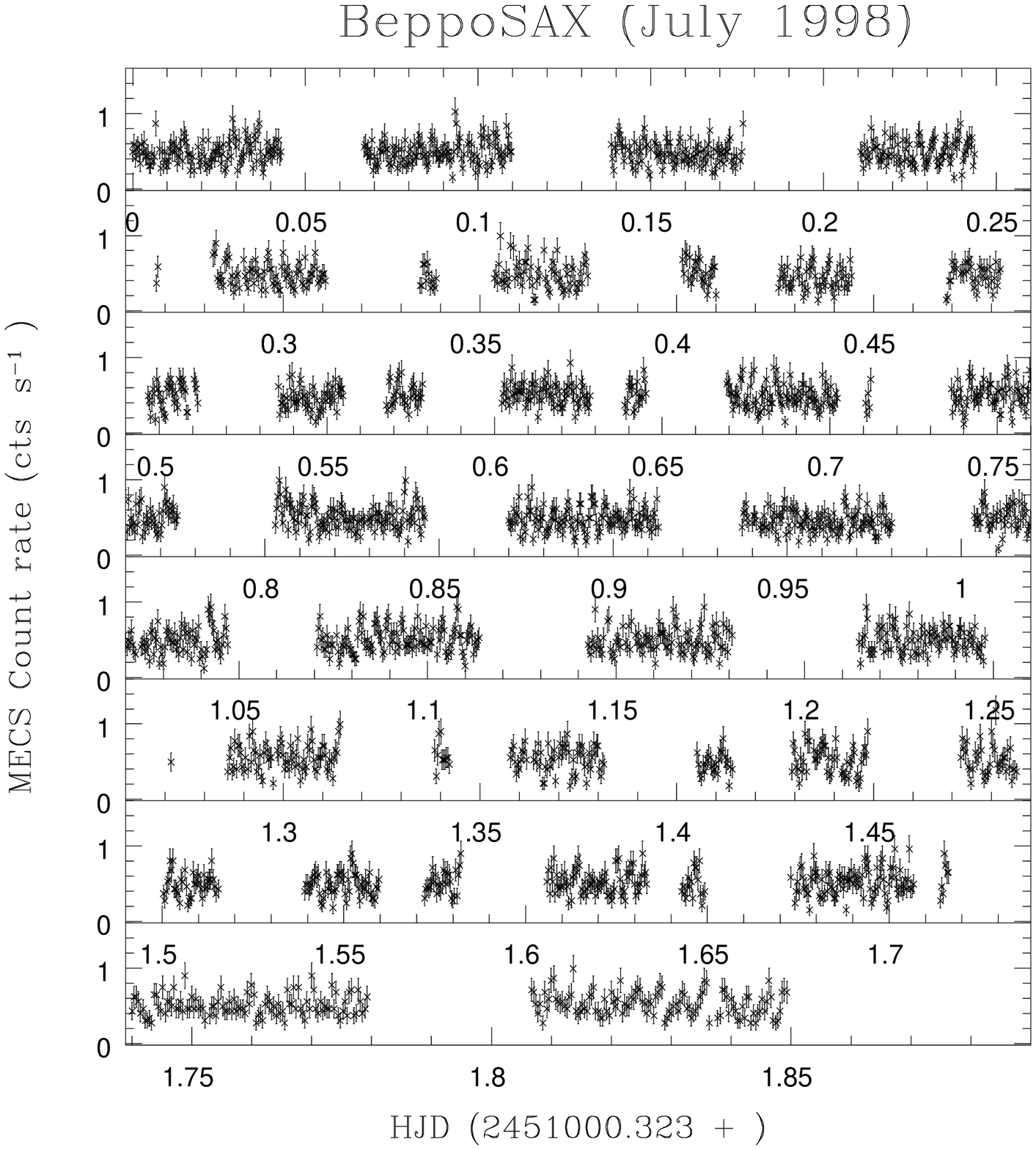}}
\caption[]{\label{curves} The net  RXTE PCA (2.9-9.8\,keV) (left
panel) and  BeppoSAX MECS (1.3-10\,keV) (right
panel) 32\,s binned light
curves of V\,709 Cas in March 1997 and July 1998 respectively.}
\end{center}
\end{figure*}

\noindent Fourier analysis has been performed on PCA 4\,s and MECS 5\,s
binned light
curves using the DFT algorithm (Deeming 1975). The corresponding 
power spectra, shown in Fig.\,2 (lower panels), reveal a strong peak at
the known 312.8\,s period at both epochs. While the 
 first and second harmonics are observed in the MECS spectrum, the PCA
spectrum lacks power at the first harmonic.  
In order to remove the windowing effects of
unevenly sampled data (see spectral windows inserted in the 
lower panels of Fig.\,2), the CLEAN algorithm
(Roberts et al. 1987) has been used, adopting a gain of 0.1 and 500
iterations. From the CLEANed power spectra, shown in Fig.\,2 (upper
panels), neither low frequency (orbital) periodicity nor  
the beat or other orbital sidebands can be detected.

\begin{figure*}
\begin{center}
\mbox{\epsfxsize=9cm\epsfbox{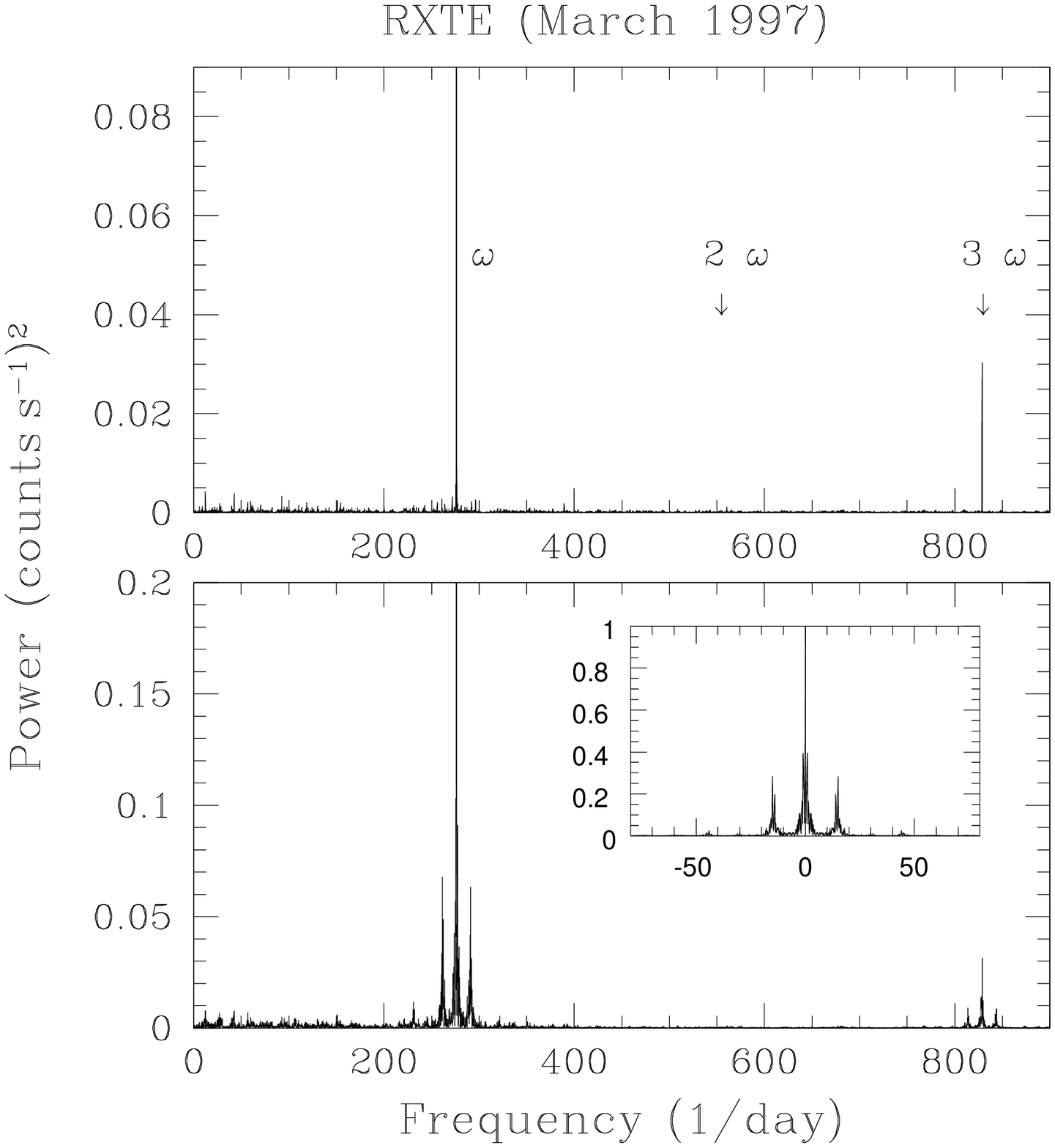}\epsfxsize=9cm\epsfbox{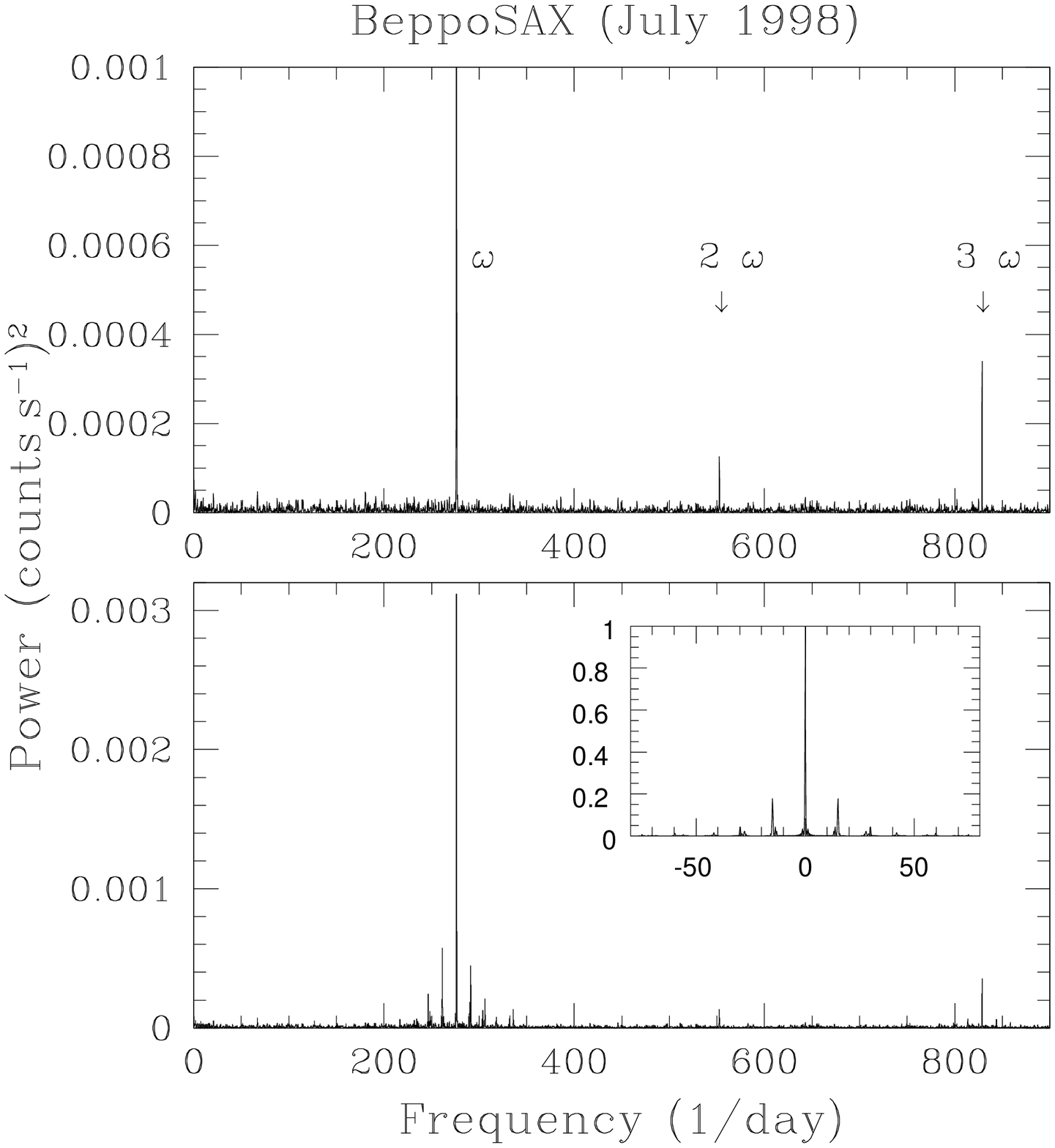}}
\caption[]{\label{mecsdft} The PCA (left panel) and the MECS (right
panel) power spectra of 
V\,709 Cas. {\it Lower panels:} The DFTs show a
strong signal at  276.26\,d$^{-1}$. In July 1998
both first and second 
harmonics are present, while in March 1997 the first harmonic is not
detected. The spectral windows are shown in the inserted panels. {\it
Upper
panels:} The removal of windowing effects is shown 
in the CLEANed spectra.}
\end{center}
\end{figure*}

\noindent From the MECS and PCA CLEANed spectra the amplitude of the main
signal is $\rm A_{\omega} = 0.108\pm 0.001\, cts\,s^{-1}$ and 
$\rm A_{\omega} = 0.82\pm 0.01\, cts\,s^{-1}$ respectively, 
corresponding to a fractional amplitude of 23$\%$ and 18$\%$ at the two
epochs. A larger amplitude in the July 1998 BeppoSAX data
is also found for the second harmonic, $\rm A_{3\omega}=0.034 \pm
0.002\, cts\,s^{-1}$ (fractional amplitude $7.5\%$), with respect to the
March 1997 data, $\rm A_{3\omega}=0.33\pm 0.01\, cts\,s^{-1}$ (fractional
amplitude of $6.5\%$). This also translates in the lack of detection
of the first harmonic in the RXTE data while in the BeppoSAX observations
the first harmonic is definitively detected, with an amplitude of $\rm
A_{2\omega}=0.019 \pm 0.002\, cts\,s^{-1}$
(fractional amplitude $3.8\%$). This fullfils the  4$\sigma$
criterium, with $\sigma=4.3\times 10^{-3} \rm cts\,s^{-1}$ being the
average amplitude from the residual CLEANed spectrum. Here we note that
the ROSAT HRI
observation carried out in February 1998 (Norton et al. 1999), and hence
 about four months earlier than the BeppoSAX pointing,
also shows the presence of the
first and second harmonics. However, the ratios of the power at
the fundamental frequency to the  first and to the second
harmonics in the BeppoSAX
LECS (0.1--2.4\,keV) range are about 90 and 10 respectively , a factor of
three and two  larger than those derived from the ROSAT HRI observations.

\begin{figure*}[t]
\begin{center}
\mbox{\epsfxsize=9cm\epsfbox{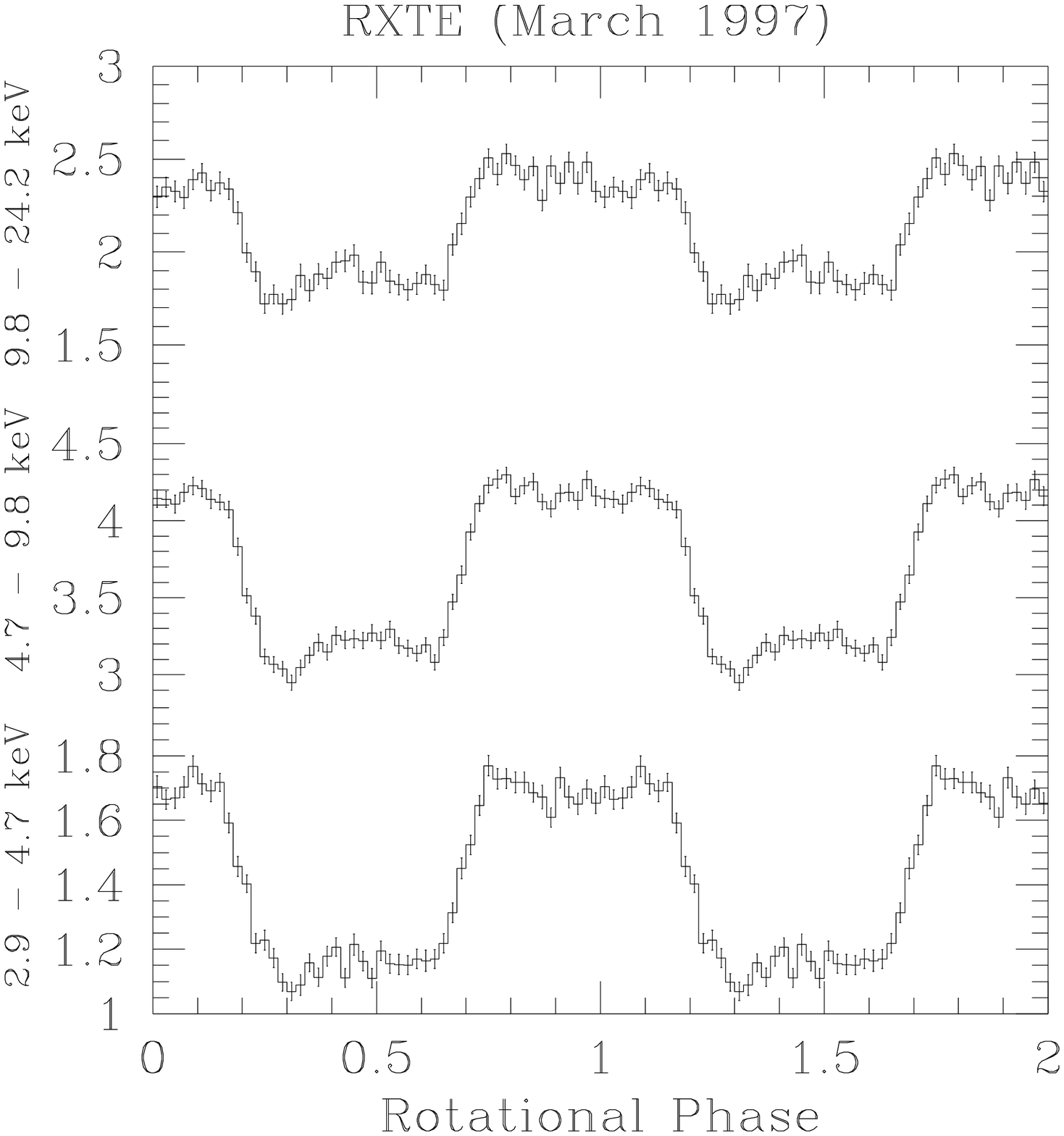}\epsfxsize=9cm\epsfbox{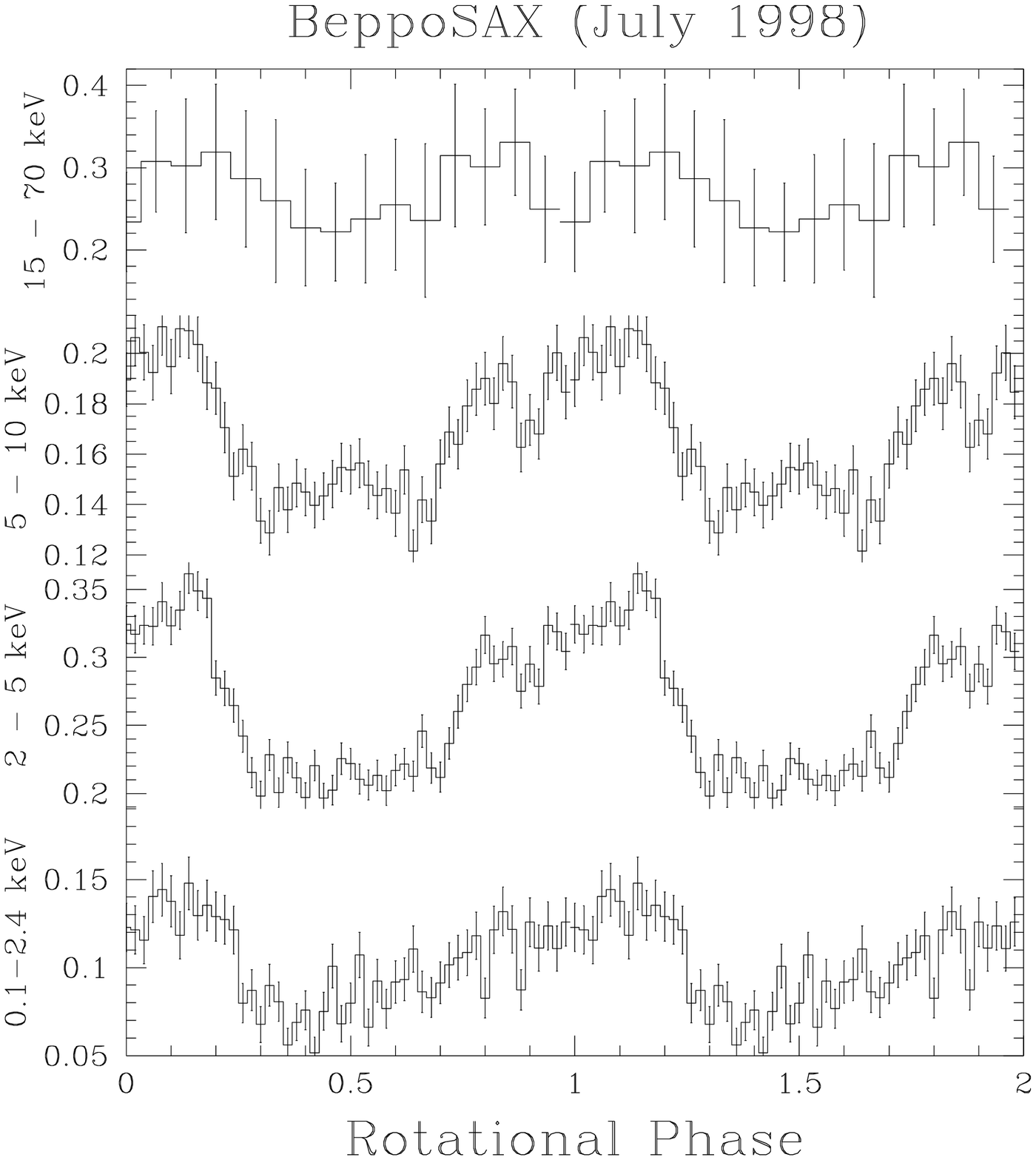}}
\caption[]{\label{foldlc} {\it Left panel:} The  RXTE light curves 
in the 2.9--4.7\,keV, 4.7--9.8\,keV and 9.8--24.2\,keV ranges (from bottom
to top) folded along the rotational period using the ephemeris reported
in the text. {\it Right panel:} The BeppoSAX 
net spin light curves in the
energy ranges (from bottom to top):
0.1-2.4\,keV, 2-5\,keV, 5-10\,keV and 15-70\,keV.
 A 6.26\,s bin size has been used except for the PDS (15-70\,keV)
light curve for which a 21\,s bin has been adopted.}
\end{center}
\end{figure*}

\noindent A sinusoidal fit to the MECS light curve using three  
frequencies  corresponding to the  fundamental, first and second harmonics
yields to a period of 312.77 $\pm$ 0.01\,s. Instead, for the PCA light
curve two sinusoids, corresponding to the fundamental and second
harmonic, have been used, which give $\rm P_{\omega}=312.746\pm0.003\,s $.
The present determinations are  compatible within errors
with the previously determined 312.78$\pm$0.03\,s period from the ROSAT
data (Haberl \& Motch 1995; Norton et al. 1999).
\noindent Times of maxima for each data set are
derived: {\it $HJD_{max}$= 2450533.647454(3)} and {\it $HJD_{max}$=
2451000.32346(2)}. The  accuracy of the derived ephemeris, however,
does not allow to bridge the gap between  the RXTE and 
BeppoSAX data sets.
 
\noindent The shape of rotational pulse has been inspected in different
energy
bands (Fig.\,3). The RXTE data in the 2.9--4.7\,keV, 4.7--9.8\,keV
and 9.8--24.2\,keV bands (Fig.\,3, left panel) have been folded
in 50 bins along the rotational period using the corresponding time of
maximum. The BeppoSAX data instead have been extracted in 
 the 0.1-2.4\,keV LECS
range (for a direct comparison with the ROSAT light curve see
Fig.4), in the
2-5\,keV and 5-10\,keV MECS ranges as well as in the 15-70\,keV PDS band.
The spin light curves, also folded in 50 bins
 (except for the PDS where a larger bin size (21\,s) has
been used), are shown in Fig.\,3 (right panel). 
At  both epochs, the rotational pulse is not
sinusoidal and shows a broad structured maximum 
(extending 0.46 in phase).  The spin pulse as
observed by RXTE has relatively symmetrical rise and decay. On the 
other hand, the  BeppoSAX light curve is highly asymmetrical with a 
slow rise and a steeper decay.  The maximum is not flat with count rates
increasing until $\phi=0.12$.  A dip feature with fractional depth
$\sim 8\%$ can be 
recognized in the hard MECS bands at $\phi$=0.9. This
feature does not occur at mid-maximum and it is less pronounced in the
soft band.  This feature could also be present in the RXTE
light curve, but its depth is only marginally significant. The changes in
the
shape of the spin modulation between
the two epochs then explain the presence of
the first harmonic in the BeppoSAX data and the lack of such 
component in the RXTE data.

\begin{figure}
%\begin{center}
\epsfig{file=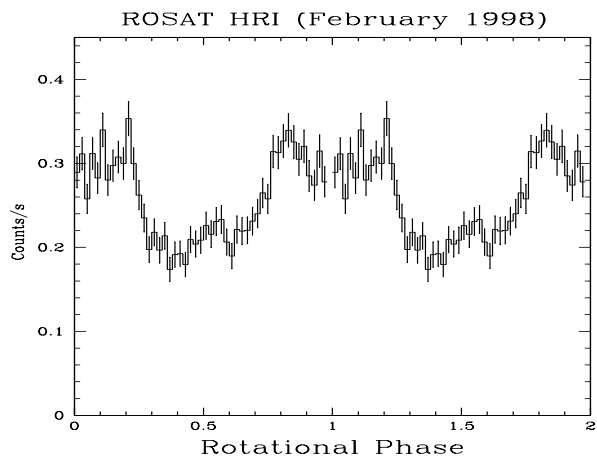, height=6.cm, width=9.cm}
\caption[]{\label{rosat}  The ROSAT HRI rotational pulsation in the
0.1--2.4\,keV band, observed by Norton et al. (1999) has been folded at
the period of 312.77\,s adopting a bin size of 6.26\,s. Phase
zero is arbitrary (see text). 
}
%\end{center}
\end{figure}

\begin{figure*}
\begin{center}
\mbox{\epsfxsize=9cm\epsfbox{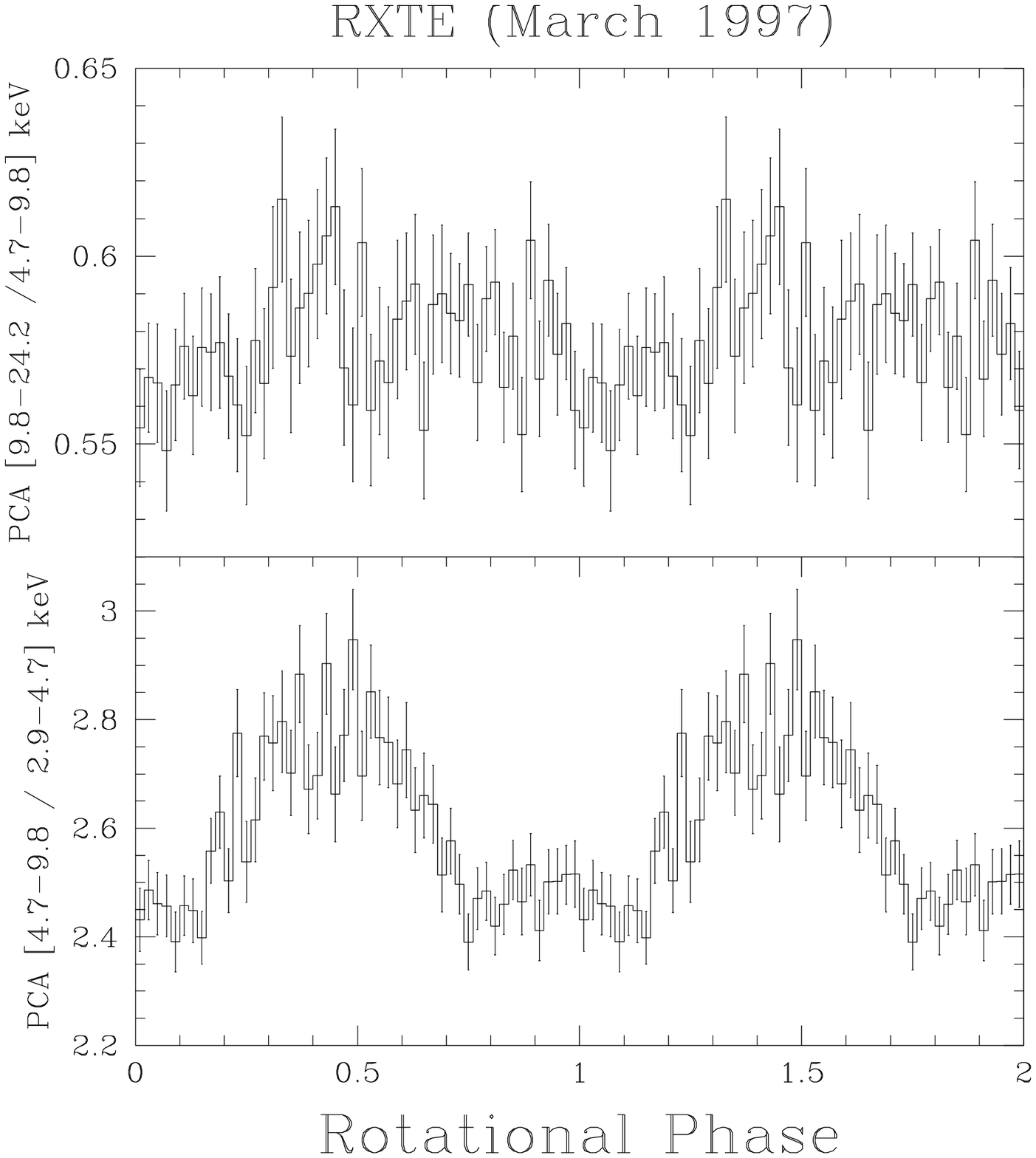}\epsfxsize=9cm\epsfbox{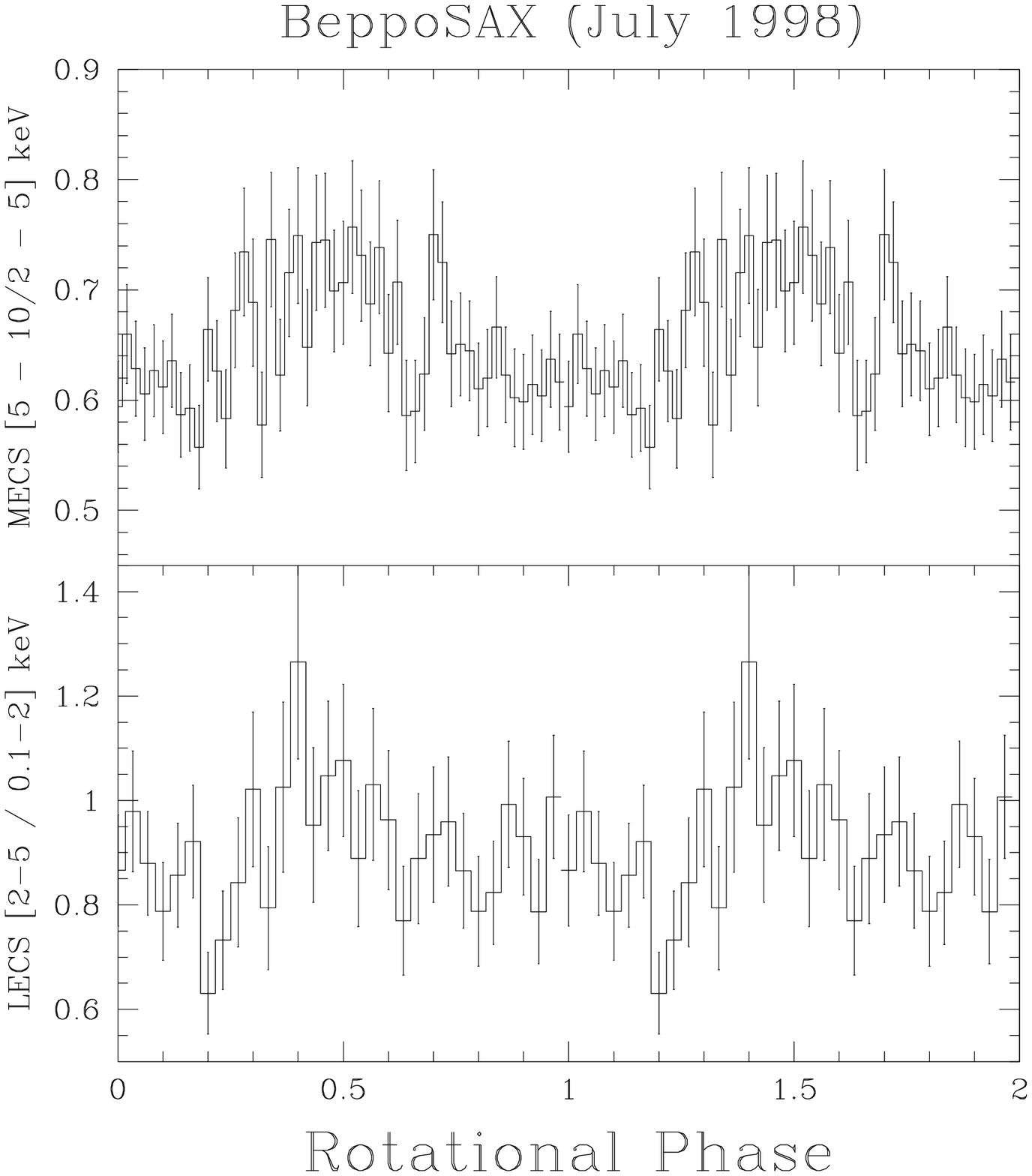}}
\caption[]{\label{foldlc} The behaviour of hardness ratios in the soft and
hard X-ray bands along the rotational period. {\it Left panel:}
the RXTE PCA 5--10\,keV and 3--5\,keV ratios (lower panel) and the
10--24\,keV and 5--10\,keV ratios (upper panel). {\it Right panel:} the
BeppoSAX MECS 5--10\,keV
and 2--5\,keV ratios (upper panel) and the LECS
2--5\,keV and 0.1--2\,keV flux ratios (lower panel). Both show a spectral
hardening at rotational minimum. The effect decreases at increasing
energies.}
\end{center}
\end{figure*}

\noindent The spin pulsation decreases in amplitude from low to high
energies:
the pulsed fraction in the RXTE data changes from 46$\%$ in the
2.9--4.7\,keV range to 32$\%$ in the 9.8--24.2\,keV band,
and in the BeppoSAX data  it decreases
from 86$\%$ in the 0.1--2.4\,keV, to 56$\%$ in the 2--5\,keV and to
52$\%$ in the 5--10\,keV band. In the PDS range, the variability is not
statistically significant.  The light curves reveal an 
energy dependence of the pulse
shape changing from a strongly asymmetrical (at low energies)  to a broad 
(at high energies) structured shape. A comparison of the low
energy pulsation in the 0.1--2.4\,keV (ROSAT--like) band  observed in July
1998 with the February 1998 ROSAT HRI rotational curve (Norton et
al. 1999) shows  a different
morphology.  The ROSAT spin pulsations is reported in Fig.\,4 with an
arbitrary phasing since the BeppoSAX ephemeris is not accurate enough.
The fractional amplitude of
the ROSAT spin pulse is about half that observed by BeppoSAX. Also,  the
broad maximum in the BeppoSAX data still
covers the same phase range but the "double-peaked" shape seen in
the ROSAT data, with a secondary minimum
centred on the broad maximum,  has essentially disappeared.  
The dip in the high energy light curve does not seem to
have a counterpart in the soft X-ray pulse. This feature is
not centered on the maximum as observed in the ROSAT light
curve.  Although a direct comparison of the RXTE and ROSAT light curves
should not be performed due to the different energy coverages, 
 the flat--topped RXTE light curve resembles more closely the ROSAT spin
pulsation with two almost identical maxima.  Again, the detection of 
an increasing power of the first harmonic in the RXTE, ROSAT
and BeppoSAX observations accounts for such differences.

%The dip has a depth which increases at higher energies. Therefore
%the energy dependence of such feature can be due to
%absorption effects at the earlier phases of the broad maximum, which
%increase  at decreasing energies.
%The difference 
%in morphology at different epochs strongly indicates  
%a long term variability in the spin pulse characteristics.

\noindent The energy dependence of the rotational pulse manifests itself
in the hardness
ratios in both soft and hard X-ray bands. In Fig.\,5 (right panel), the
LECS 
(2-5\,keV/0.1-2\,keV) and MECS (5-10\,keV/2-5\,keV) ratios show a
hardening at rotational minimum and a softening at rotational maximum. 
This behaviour is also observed in the RXTE PCA data  (Fig.\,5, left
panel), which also shows a negligible energy dependence of spin pulses
above 10\,keV.  A hardening at pulsation minimum is 
a typical behaviour in  IPs, which is produced by the larger  
photoelectric absorption when viewing along the accretion curtain. We
note that the BeppoSAX hardness
ratios do not reveal changes during the maximum phases, which would be
expected  if the double-humped maximum is due to enhanced absorption
effects within one accretion column. As it will be discussed later, the 
presence of two accreting poles can account for this behaviour.
At higher energies a different mechanism responsible for the spin
pulsation could be at work as it will be discussed in sect.\,4 and 5.

\section{Spectral Analysis}

An analysis of the time--averaged X-ray  spectra of both RXTE and
BeppoSAX observations has been performed in order to estimate
the basic parameters of the emitting region, in particular the
temperature of the optically thin plasma, interstellar and 
possible circumstellar  absorptions  and to obtain information on 
the reprocessing of the hard X-rays from the WD
surface. The spectral analysis
has been performed with the XSPEC fitting package. Henceforth, all
quoted errors on spectral parameters refer to 90$\%$ confidence level for
 the parameter of  interest.
The BeppoSAX spectra have been fitted between
0.3--70\,keV, while the RXTE ones between 3.6--20\,keV.

\begin{figure*}[t]
%\begin{center}
\epsfig{file=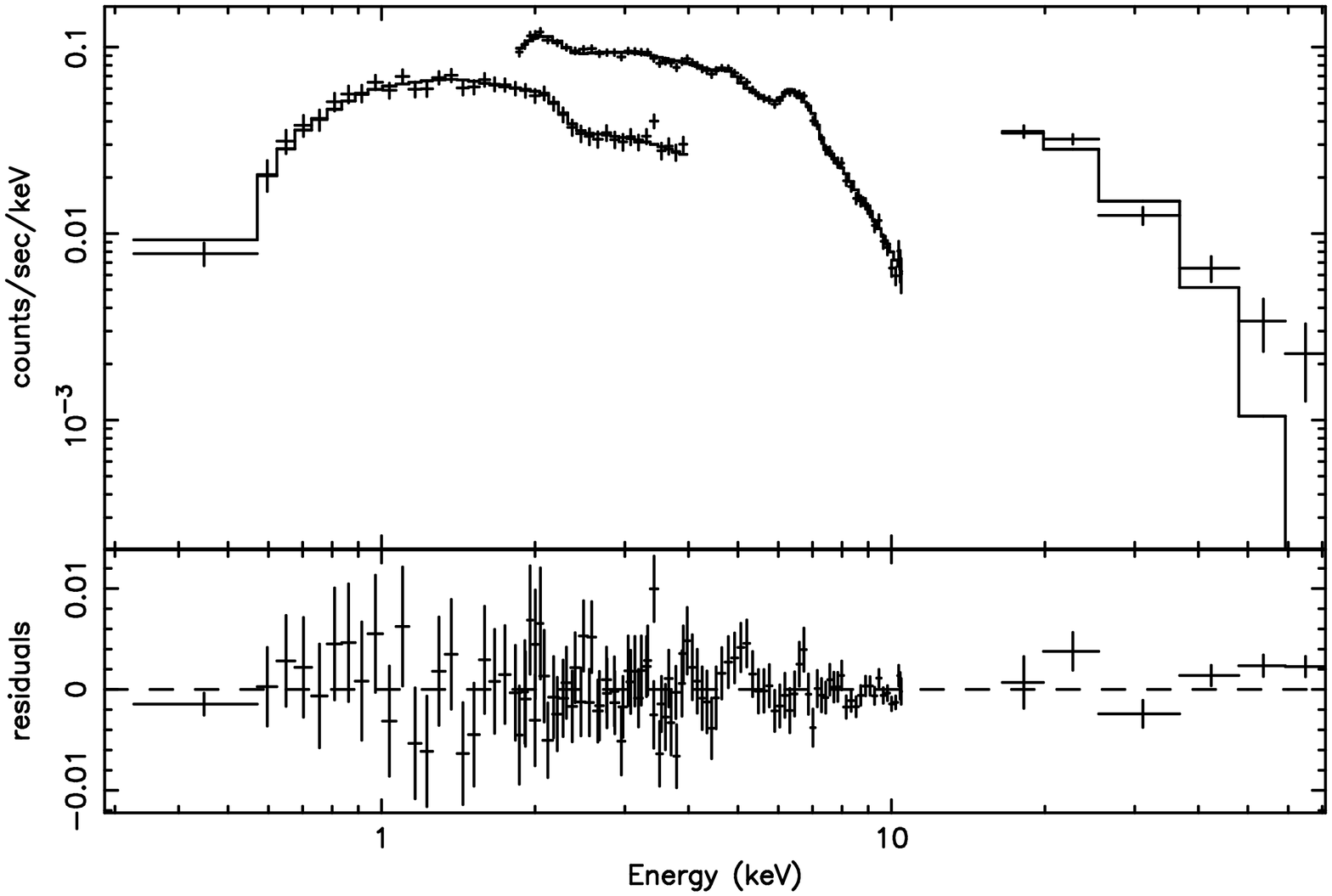, height=10.cm, width=8.cm, angle=-90}
\epsfig{file=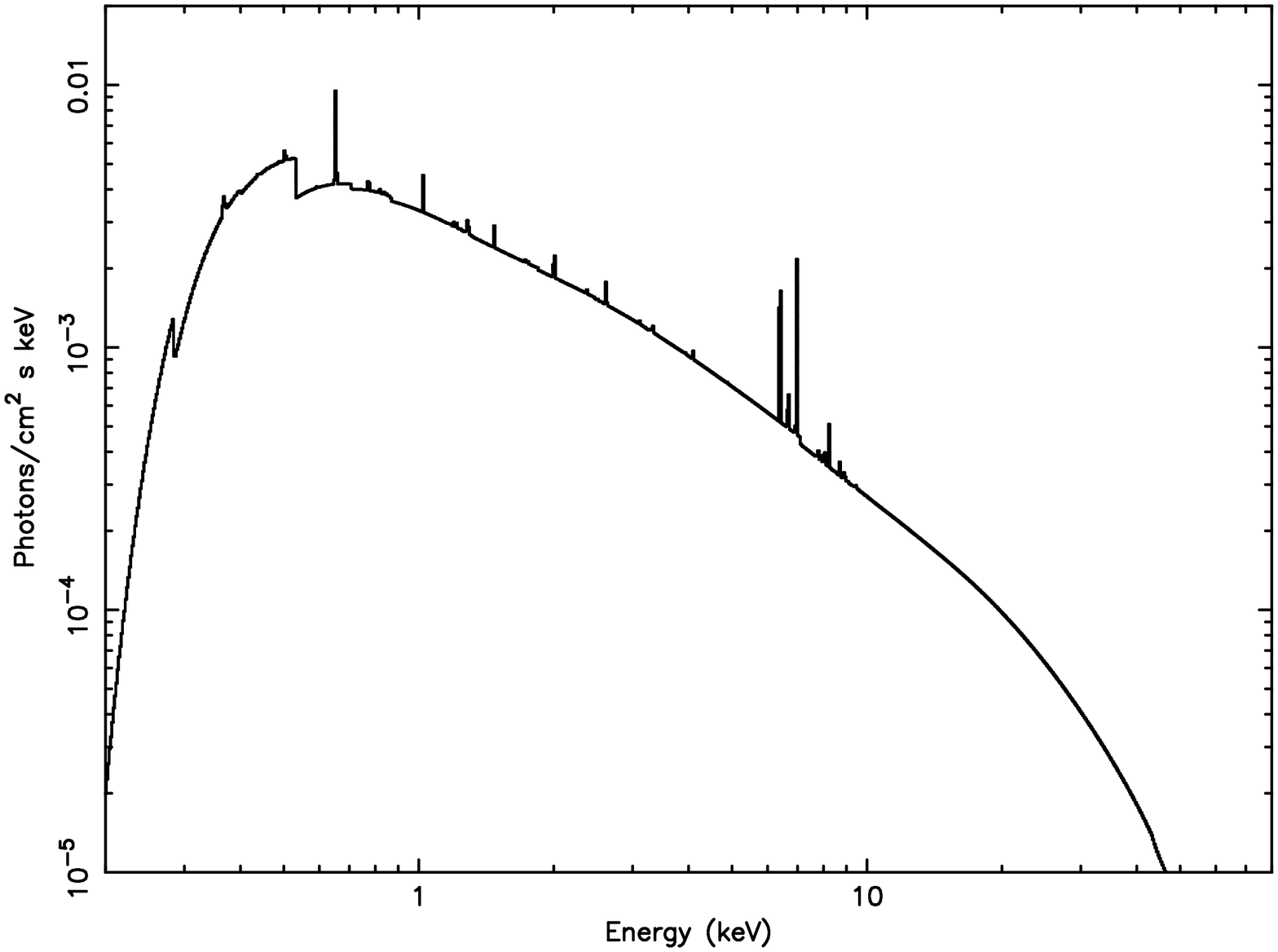, height=9.3cm, width=8.cm,angle=-90}
\caption[]{\label{average} {\it Left Panel:} The grand average
BeppoSAX LECS, MECS and PDS spectra fitted with an optically thin
plasma at kT=27\,keV together with an interstellar and partial
covering absorbers and a Compton reflected continuum and Gaussian
centred at 6.4\,keV (see text).  {\it Right Panel:} The X-ray spectrum
corresponding to the composite model is displayed in physical units.}
%\end{center}
\end{figure*}

\subsection{The BeppoSAX spectra}

\noindent A spectral fit consisting of
an optically thin plasma  (MEKAL model (Mewe et al. 1995) in XSPEC) with a single component absorber
does not produce a satisfactory fit ($\chi^2$/dof=243/114), the 
temperature being as high as 60\,keV. 
The hydrogen column density is $2.6\times 10^{21}\rm \,cm^{-2}$ which is
lower than the total galactic hydrogen 
column density (4.5$\times \rm 10^{21}\,cm^{-2}$) in the direction of the
source.  The residuals show an excess of counts 
in the low energy portion of the spectrum, likely due 
to a more complex absorption, as well
as around $\sim$6\,keV. 
Hence we have included a Gaussian  $\rm K_{\alpha}$ irone line
centred at 6.4\,keV  and  an  additional
(partial) absorber. The presence of this absorber is also
suggested by the energy dependence of the 
spin pulse and the strong (see below) iron line. We then obtain a
much better fit  ($\chi^2_{\nu}$ = 1.17) (see Table\,2, model 1). A
lower temperature is  derived, namely kT = 42\,keV which reflects the
effects of a complex model fitting.
Here we  also note that
a multi--temperature plasma model  (CEMEKL in XSPEC, which is built
from the MEKAL code, adopting a power-law temperature profile (Singh et 
al. 1996)) gives a
worse
fit  ($\chi^2_{\nu}$ =1.5), especially for the ionized iron lines and
the low energy portion of the spectrum. 
The hydrogen column density,
$\rm N_{H}= 8.7\times 10^{20}\,cm^{-2}$, is close to that
derived from ROSAT data ($\rm N_{H}=9.8 \times \rm
10^{20}\,cm^{-2}$ for a fixed  temperature of kT=10\,keV, Haberl \&
Motch 1995). This can be regarded as an upper limit to the
interstellar column density. For the partial absorber we find 
$\rm N_{H}=2.92\times 10^{22}\, cm^{-2}$ and a
covering fraction $\rm C_{F}=0.29$. The equivalent
width of the 6.4\,keV fluorescent line is E.W.=218\,eV. 
The metal abundance, left free to vary, is just
consistent with the solar value. Given the high temperature of the 
single temperature plasma model, 
the metal abundance is only sensititive to the iron abundance. We have
 adopted a relative iron
abundance by number of 4.7$\times10^{-5}$ (Anders \& Grevesse 1989).
The observed flux in the 2--10\,keV
range is 4.5$\times \rm 10^{-11}\, erg\, cm^{-2}\,s^{-1}$, a factor 2.5
higher than that extrapolated from the ROSAT 1992 PSPC data, adopting
kT=10\,keV, and it is similar to other measurements by
previous satellites (Motch et al. 1996). 

%The bolometric flux is
%1.34$\times 10^{-10}\, \rm erg\,cm^{-2}\,s^{-1}$.

\noindent A strong $\rm K_{\alpha}$ iron line would originate
by fluorescence either from the surface of the WD or from the
pre-shock accretion column, or both. To account for the large
equivalent
width, the column density of cold matter should be as high as $2\times \rm
10^{23}\, cm^{-2}$ (Inoue 1985), while a factor of $\sim$10 lower is found
for the partial absorber. 
Hence it seems likely that the bulk of the fluorescent line comes
from the WD surface. This suggests the presence of the
contribution of a Compton reflected continuum from the white
dwarf which is expected to go along with the fluorescent iron line 
(Matt et al. 1991; Done et al. 1992). Therefore, 
the introduction of a
Compton reflection continuum is physically plausible, and turns out to
be significant at the 99.7\% confidence level, according to the F-test
(see Table\,2, model 2).
Both temperature and fluorescent line parameters
change when adding such component: a lower
temperature (27\,keV) and E.W.=179\,eV
are derived from this fit. Hence, neglecting the
reflected contribution the temperature and the strength of the fluorescent
line are  overestimated. The grand average combined LECS, MECS and PDS
spectra
fitted with such a model are shown in Fig.\,6. The relative normalization
of the reflection component, represents the solid angle subtended by the
cold matter in units of 2$\pi$ steradians for an average value of
$cos\,\theta
= 0.5$ where $\theta$ is the angle between the line of sight and the 
normal to the reflecting surface.  The
value 0.92 is compatible with reflection of photons
from the WD  surface which subtends an angle of about 2$\pi$. 
The iron and metal abundances (linked together
in this fit) are within errors consistent with solar values. 
Furthermore the partial absorber covering fraction
and column density are, as expected, the same as in the 
previous fit without reflection.

\begin{table*}
\centering
\caption{ Best fit parameters for the phase--averaged X-ray spectra 
of V\,709 Cas}

%\small
\vspace{0.05in}
\begin{tabular}{lccccccccccc}
\hline
\hline
~ & ~ & ~ & ~ & ~ & ~ & ~ & ~ & ~ & ~ & ~ & ~  \cr
Data & \# & N$_{\rm H}$ & N$_{\rm H}$$^a$ &  $\rm C_{F}^b$ & $kT$$^c$ &
$A_Z$$^d$ & E.W.$^e$ & $\rm R_{ref}^f$ & $\tau$ & $\rm E_{o}$ & 
$\chi^2$/d.o.f.  ($\chi^2_{\nu}$)   \cr
~ & ~ & (10$^{20}$ cm$^{-2}$) & (10$^{22}$ cm$^{-2}$) & ~ & (keV) &
  & (eV)&  & ~ & (keV) &  ~ \cr
~ & ~ & ~ & ~ & ~ & ~ & ~ & ~ & ~ & ~ & ~ & ~  \cr
\noalign {\hrule}
~ & ~ & ~ & ~ & ~ & ~ & ~ & ~ & ~ & ~ & ~ & ~ \cr
SAX & 1 & 8.7$^{+7.5}_{-2.9}$ & 2.9$^{+1.6}_{-0.9}$ &
0.29$^{+0.06}_{-0.05}$ 
& 42$^{+6}_{-5}$ & 1.29$^{+0.35}_{-0.29}$ & 218$^{+22}_{-43}$ & ~ &
~ & ~ & 131/111 (1.18) \cr
SAX & 2 & 8.0$^{+3.0}_{-2.8}$ & 2.3$^{+0.9}_{-0.7}$ &
0.30$^{+0.03}_{-0.03}$
& 27$^{+6}_{-4}$ & 0.73$^{+0.28}_{-0.20}$  & 179$^{+25}_{-45}$ &
0.9$^{+0.4}_{-0.5}$ & ~ & ~ & 121/110 (1.10) \cr
~ & ~ & ~ & ~ & ~ & ~ & ~ & ~ & ~ & ~ & ~ & ~  \cr
\noalign {\hrule}
~ & ~ & ~ & ~ & ~ & ~ & ~ & ~ & ~ & ~ & ~ & ~ \cr
RXTE & 3 & ~ & 1.6$^{+0.2}_{-0.1}$ & 0.99$^{+0.01}_{-0.53}$ &
36$^{+2}_{-2}$ & 0.52$^{+0.09}_{-0.08}$ & 216$^{+15}_{-16}$ &
~ & 0.13$^{+0.02}_{-0.02}$ & 8.1$^{+0.1}_{-0.2}$ & 45/37
(1.23) \cr
RXTE & 4 & ~ & 1.7$^{+0.2}_{-0.2}$ & 0.99$^{+0.01}_{-0.55}$ &
26$^{+6}_{-4}$ & 0.40$^{+0.11}_{-0.17}$ & 198$^{+16}_{-17}$ &
0.3$^{+0.2}_{-0.2}$ & 0.11$^{+0.02}_{-0.02}$ &
8.0$^{+0.2}_{-0.2}$ & 40/36 (1.11) \cr 
~ & ~ & ~ & ~ & ~ & ~ & ~ & ~ & ~ & ~ & ~ & ~  \cr
\hline
\hline
\end{tabular}
~\par
\begin{flushleft}
$^a$ Column density of the partial absorber.\par
$^b$ Covering fraction of the partial absorber.\par
$^c$ Plasma temperature.\par
$^d$ Metal abundance in units of the cosmic value (Anders \& Grevesse 
1989).\par
$^e$ Equivalent width of the 6.4 keV fluorescent iron line.\par
$^f$ Relative normalization of the reflection component (see text).
\end{flushleft}
\end{table*}

\noindent The variability in the hardness ratios is a strong evidence that
spectral changes occur at the rotational period. A spin resolved spectral
analysis  has been
performed  using only the LECS and MECS data, due to the lack of 
information on variability from PDS data. Spectra have been 
extracted in 10 rotational phase bins.
In order to estimate changes in the spectral parameters, a 
fit with a composite model consisting of
a MEKAL emission plus the interstellar column density, a partial absorber
and Gaussian line has been performed. With the exclusion of the PDS data
the temperature of the optically thin gas has been fixed at 30\,keV,
which is the value obtained from fitting the average spectrum without
the PDS data to model\,1.  We note that the fit quality does
not substantially change if kT=40\,keV is assumed. 
As a first attempt both covering fraction $\rm C_{F}$ and $\rm N_{H}$
of the partial absorber  and the Gaussian line normalization have been
left free to vary, with the
remaining parameters fixed at the values found for the phase--averaged
spectrum (model\,1). In these fits, the partial covering parameters are
less well defined, especially
$\rm N_{H}$,  which is essentially constant within
errors. Hence
fits have been performed again varying only 
$\rm C_{F}$ and the Gaussian normalization, with
$\rm N_{H}$ of the partial absorber fixed at the value found for
model\,1. The  results are shown in Fig.\,7, where the changes in the
covering fraction (a factor $\sim 1.6$) indicate an increase in
 absorption at  rotational minimum. The E.W. of the fluorescent line
also increases at
rotational minimum by a factor $\sim 2.3$.

\begin{figure}[h!]
%\begin{center}
\epsfig{file=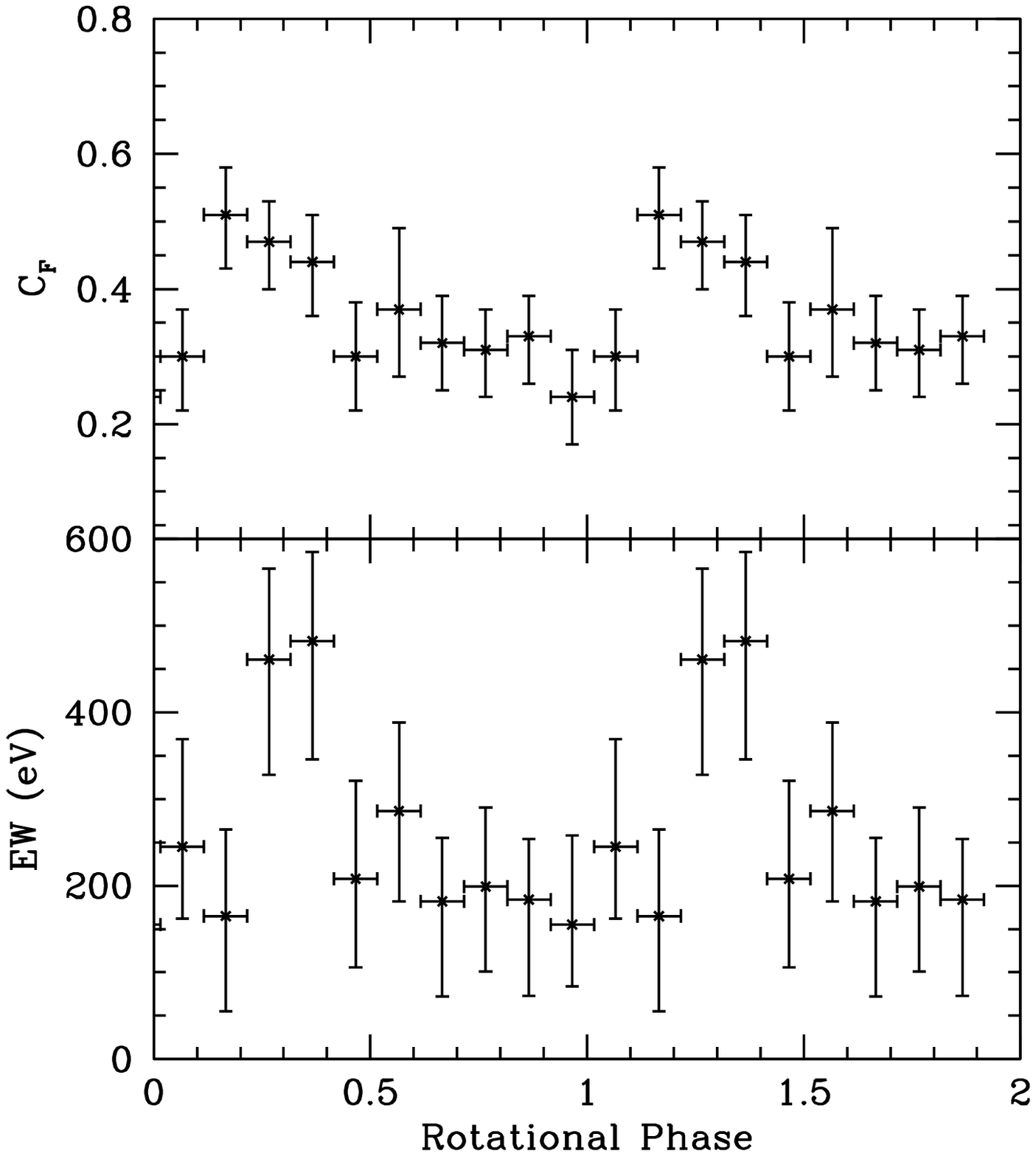, height=10.cm, width=9.cm}
\caption[]{\label{fitspin} Variations of the covering fraction of the
partial absorber and E.W. of fluorescence iron line along the spin
pulse period obtained from the BeppoSAX observation.
A composite model consisting of a MEKAL emission plus interstellar
hydrogen 
column density, partial absorber and Gaussian line has been fitted to the
combined LECS and MECS spectra extracted in 10 spin bins. The temperature,
the interstellar and partial absorber column densities and abundances have
been kept constant (see text).
}
%\end{center}
\end{figure}

\noindent An attempt to search for variations in the reflection
component has
been performed using the spectra at rotational maximum and minimum,
keeping constant the temperature, abundances, interstellar and partial
covering hydrogen column densities at the values found
from fit with model\,2, in Table\,2. The variations  of the relative
normalization
$\rm R_{refl}$ in these fits are consistent with the iron line E.W.
behaviour shown in Fig.\,7, being lower at
rotational maximum ($\rm R_{refl} \leq 0.2$) and larger at minimum ($\rm
R_{refl}=0.7\pm 0.4$). \\

\begin{figure*}[t]
%\begin{center}
\epsfig{file=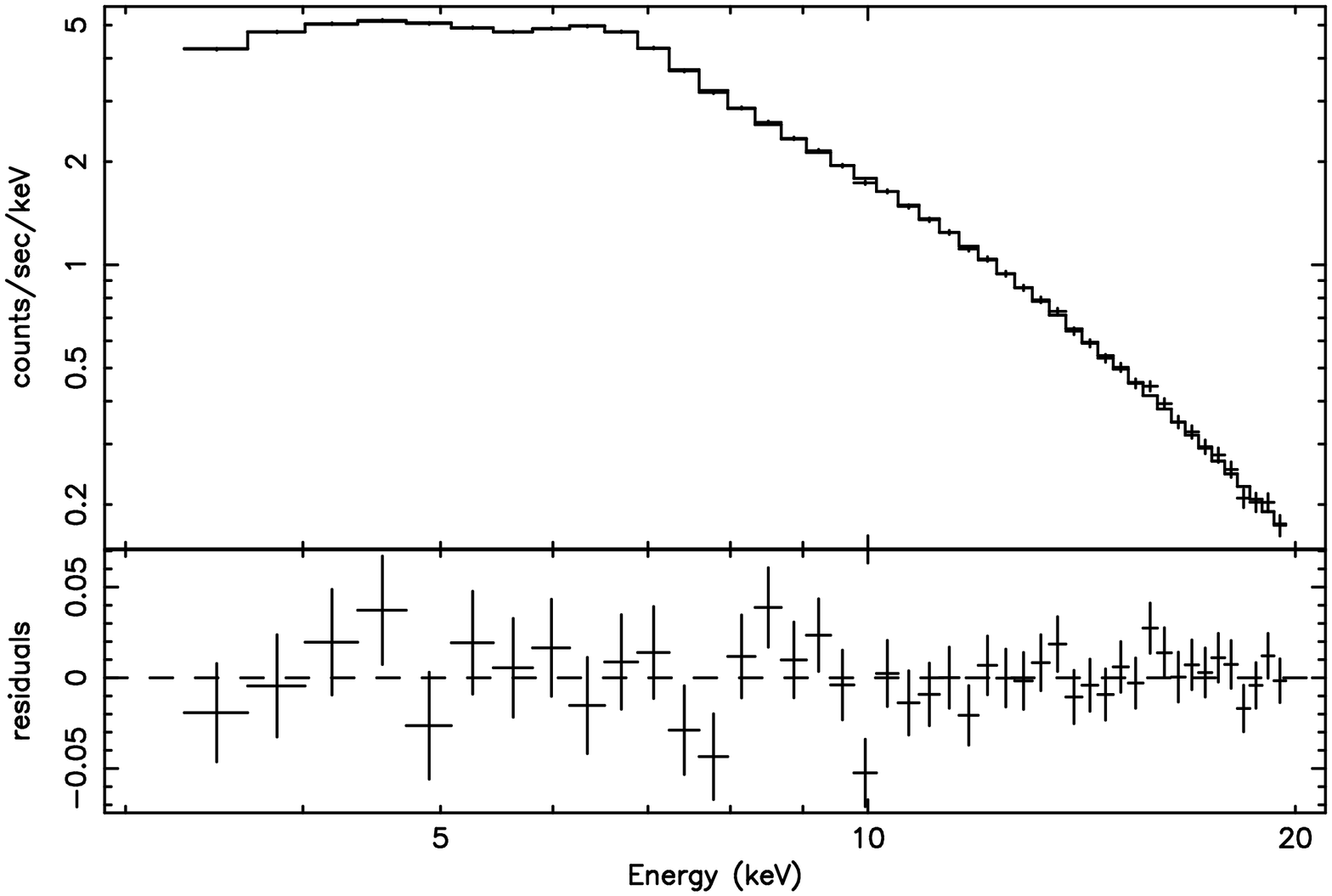, height=10.cm,width=8cm, angle=-90}
\epsfig{file=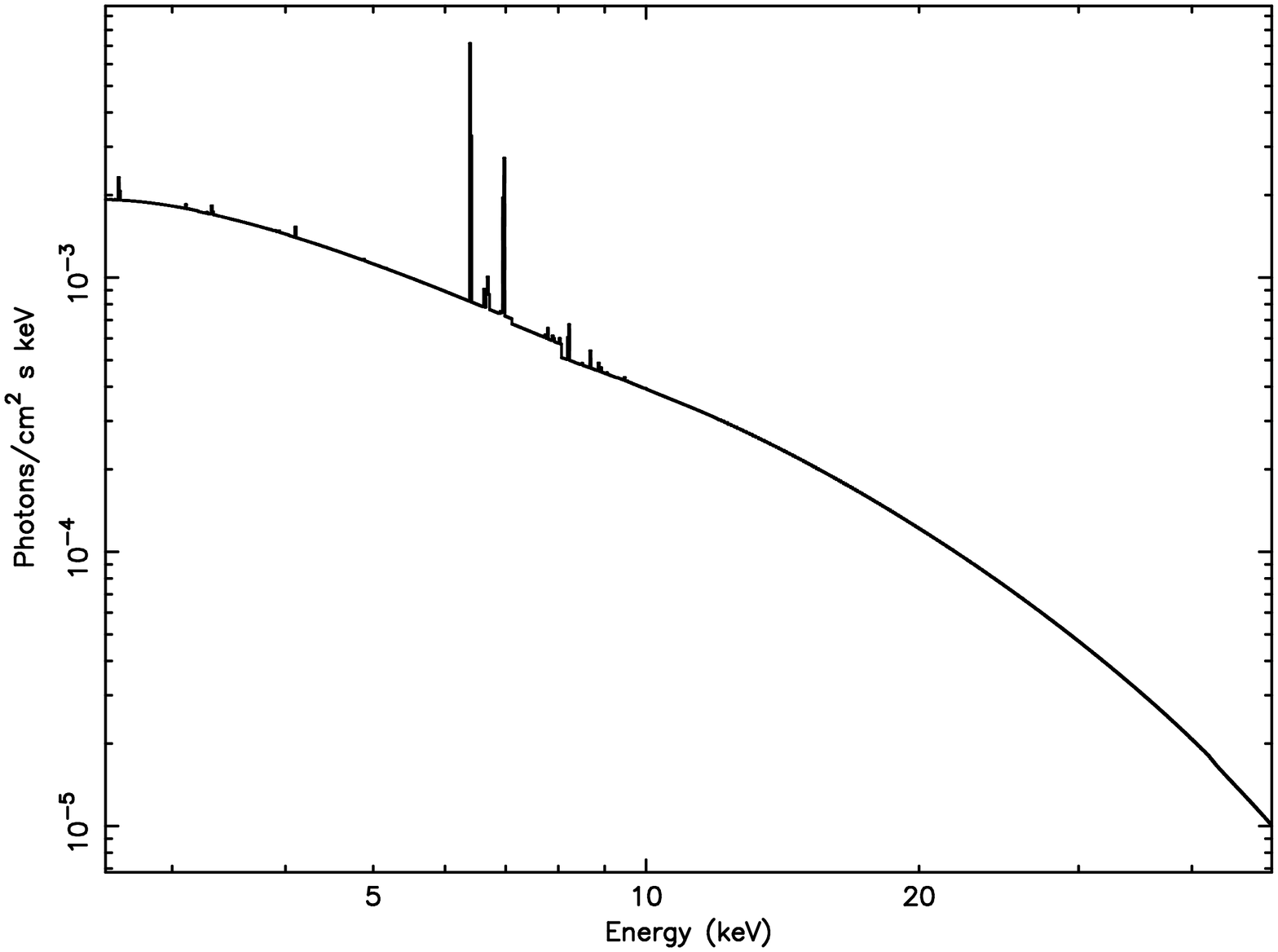, height=9.3cm,width=8cm, angle=-90}
\caption[]{\label{xtefit} {\it Left Panel:} The grand average RXTE
spectrum fitted with an optically thin plasma at kT=26\,keV together
with an absorber with $\rm N_{H}=1.7\times 10^{22}\,cm^{-2}$, an
absorption edge with $\tau=0.11$ and $\rm E_{o}=8.0\,keV$, and a
Compton reflection continuum plus a Gaussian line centred at 6.4\,keV
(model\,4).  {\it Right Panel:} The X-ray spectrum corresponding to
the composite model is displayed in physical units.}
%\end{center}
\end{figure*}

\subsection{The RXTE spectra}

The time--averaged RXTE spectrum was fitted first with a MEKAL
model, together with a Gaussian line centred at 6.4\,keV, and two
absorbers
to account for the interstellar and partial covering but the fit quality 
is bad  ($\chi^{2}_{\nu}$ = 5.97).
The residuals show a strong excess of counts around 5\,keV and a
depression around 8\,keV. The presence of this feature
has been checked against instrumental (background subtraction) effects,
and in the individual spectra of all five PCU units. We interpret this
feature as  a signature of an absorption edge.
 Furthermore the interstellar column density 
results are negligible, as expected by the limited extension to
low energies. The covering fraction of the partial absorber results
instead close to unity (total absorber). Hence the fits have
been performed again neglecting the interstellar absorption and 
introducing an additional component ($e^{-\tau(E/E_{o})^{-3}}$ at
$\rm E>E_{o}$)  in order to account for the absorption edge. As shown
in Table\,2 (model\,3), the fit 
 quality substantially improves  ($\chi^{2}_{\nu}$ =1.23). Except
for the temperature (kT=36\,keV), some 
large differences are encountered in the spectral parameters with
respect to the BeppoSAX results. First of all the presence of the
edge with $E_{o}$= 8.1\,keV and an optical depth 
$\tau = 0.13$, which is not required, although not excluded by the
BeppoSAX data. Here we note that although BeppoSAX has a better
energy resolution than RXTE, the latter  has a much larger
 collecting area. 
The edge energy corresponds, within the error, to Fe {\sc xix-xxii};
the optical depth to an equivalent hydrogen column
density of $\rm \sim 1.1\times10^{23} A_{\rm Fe}^{-1}$ cm$^{-2}$.
The abundances results seem to be much lower than
solar values. We note that fixing them to the solar value, the fit
quality decreases noticeably  ($\chi^{2}_{\nu}$=2.13). The origin of
such large difference is unclear, but it should be noted that at
the low resolution of RXTE the various iron lines are heavily blended. 
Also, the
absorber is total while column densities are within errors comparable
to the BeppoSAX fits. The presence of such  an absorber could be
consistent with the presence of the absorption edge. The equivalent
width of the $\rm K_{\alpha}$ iron
line (216\,eV) is similar to that derived from the BeppoSAX data. 
With this model the observed flux in the 2--10\,keV range is
6.64$\times 10^{-11}\, \rm erg\,cm^{-2}\,s^{-1}$, $\sim 1.4$ times
larger than that observed in July 1998 by BeppoSAX. An
attempt to fit the RXTE data with an additional reflection component
has been performed, resulting in a modest improvement in the quality
of the fit, significant at the 95.9\% confidence level (F-test; see
Table\,2 model\,4 shown in Fig.\,8). 
 The temperature is lower
(kT=26\,keV) than that derived in model\,3 and similar to that
derived from BeppoSAX data. The relative normalization of
the reflection component is  comparable, within errors,
 to the BeppoSAX fits. In summary the RXTE grand average spectrum provides
the first evidence  of absorption  from ionized material.

\noindent We also performed a phase--resolved spectral analysis of the
RXTE data,
limiting ourselves to study the maximum and minimum phases. Within
the errors, no parameter
changes between the two phases apart from the column density of the cold
absorber, which is 1.37$^{+0.23}_{-0.24}\times10^{22}$ cm$^{-2}$ at the
maximum and 2.33$^{+0.16}_{-0.34}\times10^{22}$ cm$^{-2}$ at the minimum.

\section{Discussion}

The RXTE and BeppoSAX observations have  allowed, for the first time, 
a simultaneous characterization of the X-ray temporal and spectral
 behaviour of V\,709 Cas. 

\subsection{Variability}

It has been possible to confirm the 
rotational period of the WD at 312.75\,s and to rule out
the presence of 
any orbital and sideband periodicities. This system is therefore a typical
disc accretor. The rotational modulation is remarkably
non--sinusoidal compared to most IPs. The spin pulses are clearly
observed at all energies up
to 25\,keV, with decreasing amplitudes from soft to hard X--rays, a
typical behaviour of IPs.
Furthermore, we have found that the power at the spin frequency and first
and second harmonics changes with time: the spin pulsed fraction observed
by BeppoSAX in July 1998 was twice that observed in February of the
same year by ROSAT (comparison being done  in the 0.1--2.4\,keV
range) and about 1.2 times larger than that observed by RXTE
in March 1997 ( comparing the 2-5\,keV range). Not only
the amplitude but also the
morphology of the rotational modulation has changed.
The one observed by BeppoSAX is highly asymmetrical with a maximum 
possessing two peaks with different intensities in the hard band, but it 
is single--peaked in the soft (ROSAT--like) band.  The latter
contrasts with the double--peaked maximum (with similar intensities) as
observed by ROSAT.
The BeppoSAX pulsation also differs from the very symmetrical RXTE spin 
pulse which presents a relatively flat--top maximum and 
more resembles the ROSAT light curve. It is worth noting
that the pulse maximum extends over the same phase range in all
observations. The presence of the first harmonic in the two observations
in 1998, contrasts  with the lack of detection in March 1997. 
 A strong first harmonic, compared to the second, would produce a
double peaked light curve with two maxima offsetted in phase by 180$^o$.
In V\,709 Cas the second harmonic is the stronger and produces an
asymmetrical profile when the first harmonic is present and strong, as
observed by BeppoSAX. When the first harmonic is weak or even absent
the curve possesses similar maxima as observed by ROSAT and RXTE.

\noindent The changes in the shape of light curve are also accompained by
variations in the X--ray flux.  Comparing the ROSAT PSPC count rates
observed {in  1992} (Motch et al. 1996) with those
converted from the ROSAT HRI pointing in February 1998 (Norton et
al. 1999), we find that
V\,709 Cas was brighter by a factor of $\sim 1.2$ in the
latter observation. Furthermore
converting the BeppoSAX flux in the ROSAT energy pass--band, we find that
 the binary  was brigher by a factor of
 $\sim$1.7 in July of the same year. This  suggests that V\,709 Cas
experiences changes in luminosity on time scales of
 less than months to years  and supports the identification of a
highly variable
X-ray source  made by Motch et al. (1996). Hence changes in the accretion
rate are likely to occur in this system (see sect.\,5.2) which
give rise to changes in the amplitude and shape of spin pulses. This phenomenon has
also been  observed in other IPs such as FO\,Aqr (Beardmore et al. 1998;
de Martino et al. 1999), TX\,Col (Norton et al. 1997) and BG\,CMi (de
Martino et al. 1995). 
As the accretion disc  extends down to the magnetospheric radius,
which scales as $\propto \dot M^{-2/7}$, an increase in the mass
accretion rate can produce a shrinking of the magnetospheric
radius and hence an accretion spread over larger areas of the
WD surface. At higher accretion rates the effects of 
absorption are expected to be larger, thus influencing the
shapes and amplitudes of the low energy rotational modulation.
The increase in the spin amplitude between the February and
July 1998 could be consistent with this interpretation. On the other
hand, the March 1997 RXTE spin pulse has a lower amplitude with respect
to that in 1998. At highest accretion rates, some of the local
absorbers become ionized, which can reduce the low energy modulation
amplitudes.

\noindent  The double-humped light curve observed
by ROSAT was attributed to the contributions of both accreting poles 
by Norton et al. (1999), but with an offset of the dipole axis
thus producing two maxima separated by less than half a spin period.
In the two--pole accretion scenario there are two ways to produce  a
double--peaked spin pulsation. One invokes large areas, due to the 
low magnetic field of the accreting WD, with vertical optical
depth (along field lines)  lower than the horizontal optical depth.
The other  invokes both wide and tall shocks, but with the  
 vertical optical depth larger than the horizontal, as in the
classical accretion
scenario. The RXTE and BeppoSAX observations have shown that the spin 
pulse is energy dependent and the pulses are in
phase at all energies (i.e. above and below 10\,keV). To account for a
rotational modulation above 10\,keV, reflection cannot be the main
mechanism, since, as shown in  sect.\,4, it would produce an 
anti--phased pulse with respect to the low energy pulsation. A self
occultation of tall shocks was proposed as a solution 
for the high energy rotational modulation in IPs (Mukai 1999). This can 
account for the phasing of spin pulses at both high and low
energies,  for the lack of an energy dependence of the hard X-rays
pulsations and for the sharpness of the rise and decay of the pulsation.
Hence the possibility that in V\,709 Cas the two accreting
poles are wide and tall, with the optical depth along the magnetic field
lines larger than across  them (as in the classical accretion
curtainmodel), seems to be the most viable. Since
V\,709 Cas is a classical disc--accretor, the accretion is expected
to occur onto both poles. Also, in order to have the same phasing 
at both high and low energies, $i + \delta < 90^{o}$, where $i$ is 
the binary inclination and $\delta$ is the magnetic colatitude (Mukai
1999),
with an upper limit to the inclination of $i < 68^{o}$ due to the
lack of an X-ray eclipse in this system. 
This would  imply that the upper pole dominates the hard
X-rays at all phases and hence does not produce a strong modulation,
but the lower
pole produces a maximum when the upper pole points away from
the observer. The lower pole appears into view for $\sim 40\%$ of 
a cycle, thus producing the sharp rise to and sharp decay from the
maximum.  When the upper pole points towards the observer (spin minimum),
the lower pole is out of view. A simplified sketch of the geometry is 
shown in Fig.\,9 for an illustrative set of values, $i=45^o$,
$\delta=30^o$ and shock height $h = 0.2\,R_{wd}$ for two symmetrical
poles. An asymmetry between the two poles was pointed out by Norton et 
al (1999) which seems to be necessary to explain the details of the
light curves. An offset value of  $\sim 108^{o}$ could be
derived from the separation of the two maxima in the ROSAT HRI light
curve. However for a detailed modeling of the accretion
geometry one should take into account the whole three-dimensional shape of
the emission region which would imply too many degrees of freedom.
For our purposes, we only note that the net effects of the
absorption within the two intervening accretion curtains at their
respective spin maximum will then 
harden the spectrum at these phases. This explains
the lack of spectral changes during the maximum and dip phases.
Furthermore, the BeppoSAX observation shows that this maximum
is highly asymmetrical, which  would suggest
that the contribution from some parts of the
accretion regions are not identical at some epochs.

\begin{figure}[h!]
%\begin{center}
\epsfig{file=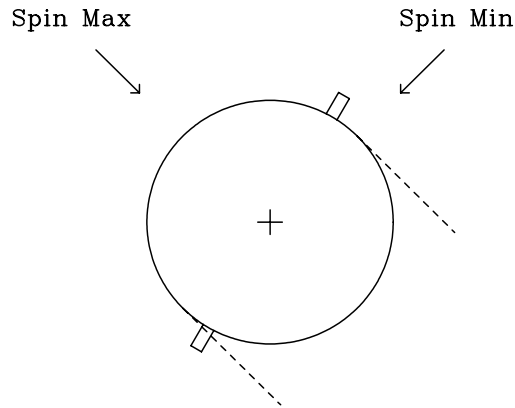, height=6.cm, width=8.cm}
\caption[]{\label{sketch} {A schematic sketch of the X-ray spin modulation
in V\,709 Cas for $i=45^o$, $\delta=30^o$ and a shock height of 0.2$\rm
R_{wd}$ assuming for semplicity two symmetrical accreting poles. At spin
minimum (the observer's line of sight is indicated by the arrow from the
upper right), the lower pole is hidden, while the upper pole is seen
along the field lines. At spin maximum (the line of sight is shown from 
the left), the entire upper pole and most of the lower pole are visible 
(the regions hidden by the WD are indicated by the dashed lines), and both
poles are seen across the field lines. The lower pole is visible for
$\sim 40\%$ of a cycle, thus being responsible for the broad maximum. The
upper pole does not produce a strong modulation, being visible at all spin
phases.}
}
%\end{center}
\end{figure}

\subsection{Spectral Properties}
 
The spectral analysis carried out from 0.1 to 100\,keV has shown that
V\,709 Cas is a classical IP  in its X-ray characteristics.
A model consisting of a single temperature optically thin emitting gas,
attenuated by both interstellar material and a partial absorber, and
 an emission feature corresponding to the fluorescent 6.4\,keV iron
line, describes well the BeppoSAX X-ray spectrum. The post-shock
temperature  ranges between 42 and  27\,keV, depending on whether 
reflection from the WD surface is
included in the spectral fits. Within errors the difference between the
two temperatures is not significant.  We  also
note that a multi--temperature plasma
is not required by either RXTE  or BeppoSAX data. In this
respect, we compare
the derived temperature of the post--shock region with that 
expected from the simple shock model

$\rm kT_{shock} =3/8\,G\,\mu\,m_{H}\,M_{wd}\,R_{wd}^{-1}$ 

\noindent with $\mu$ the mean molecular weight and $\rm m_{H}$ the mass of
hydrogen. From optical spectroscopy, Bonnet-Bidaud et
al. (2001) constrain the WD radius at
a value $\sim 8.7\times 10^{8}$\,cm and a mass of 0.75$\rm M_{\odot}$,
which then implies a temperature of 30\,keV, in close agreement with
 our BeppoSAX and RXTE results.

%We note that multi--temperature
%models work well only if there is a lower limit of a few keV to the
%temperature as in the case of V\,1223 Sgr (Beardmore et al. 2000).

\noindent From the spectral fits we have found that  the column
density of the partial absorber cannot  alone explain the large
equivalent width  ($\sim$ 200\,eV) of the 6.4\,keV line. Hence the
contribution
from reflection  by the cold material of the WD surface should 
dominate the fluorescent iron line. At a temperature of $\sim$30\,keV,
an EW  $\ge$ 100\,eV is expected  for $\theta
<36^{o}$ (Matt 1999), where $\theta$ is the viewing angle between the
line of sight and the direction orthogonal to the illuminated portion 
of the WD surface.  From Fig.\,9, the largest
contribution to the iron line from the reflected region of the WD surface
should then be produced by the upper pole at spin minimum ($\theta
=15^o$). However if some asymmetry exists, this enables to see
the lower pole more  effectively, thus giving rise to a large equivalent
width also at other spin phases.
%Hence at all spin phases the viewing angle is
%constrained, giving  $cos\,i\,\,cos\,\delta -
%sin\,i\,\,sin\,\delta\,\,cos\,\phi >$ 0.8, where $\phi$ is the phase. 
%
\noindent We have  indeed found spectral variations  with the
rotational period, which indicate a spin
variability of the partial absorber and emission line parameters,
the largest contribution of both being at rotational minimum. Again,
the accretion curtain scenario matches well the spectral behaviour of
spin pulses. When the curtain points towards the observer, the absorption
effects are larger (the covering factor is larger) as well as the effects
produced by reflection, since the accretion area projected on the WD 
is larger. 
This is also seen from fitting the spectra at spin maximum and minimum
which show an increase in the Compton reflection component at spin
minimum.

\noindent The spectral analysis of 1997 RXTE data has furthermore 
shown the presence of an absorption feature at $\sim$ 8.1\,keV 
which is interpreted as an absorption edge from  Fe {\sc xix--xxii} with
an optical depth $\tau \sim 0.11-0.13$. This would
imply an ionization parameter of $\sim$ 100 (Kallman \& Mc\,Cray
1982). This is the first detection of an ionized
absorber in an IP. Such material could be located in the pre--shock
regions of the accretion funnel. The lack of detection of a
variability of this feature at the rotational period might 
suggest that this material is distributed over
a large angle as seen by the observer. The results of the RXTE spectral
fits also suggest the presence of a total absorber, which is consistent
with a large covering area.

\noindent We attempt to obtain estimates of the accretion rate from 
the X--ray luminosities of BeppoSAX and RXTE observations.
 Bonnet-Bidaud et al (2001) constrain V\,709 Cas at a distance
between 210--250\,pc, which would imply a bolometric
luminosity of 1.1--1.5 $\rm \times 10^{33}\,erg\,s^{-1}$ during the 
March 1997 RXTE observations and of 0.8--1.1$\rm \times
10^{33}\,erg\,s^{-1}$ during the July 1998 BeppoSAX observations.
This gives an accretion rate of 9.6--13.0$\rm \times 10^{15}\,g\,s^{-1}$
in 1997 and of 6.9--9.6$\rm \times 10^{15}\,g\,s^{-1}$ in 1998. 
The accretion rate has then decreased by a factor of $\sim$1.4
between 1997 and 1998. Hence the detection of an ionized material
in the RXTE data could be consistent with a higher
accretion rate experienced in 1997.

\noindent We also derive an average emission measure 
E.M.=2.6--3.8 $\rm \times 10^{55}\,cm^{-3}$ and E.M.= 1.7--2.4 $\rm
\times
10^{55}\,cm^{-3}$ for the March 1997 and July 1998 observations, assuming
the above range of distances. As EM $\sim n_{H}^{2} \times l^{3}$, we give
an upper limit to the number density
of hydrogen and an estimate to the linear dimension of the post--shock
region.
We use the RXTE results since from the
iron absorption edge, we have derived the equivalent hydrogen column 
density $\rm N_{H} \sim 1.1\times 10^{23}\,
A_{Fe}^{-1}\,cm^{-2}$. Assuming that this
material is in the pre-shock region, and for simplicity in the
immediate vicinity of the shock, the density in the region just
below the shock is a factor of four higher.  Then, from 
the range of emission measure, we 
obtain $l\sim 3.4-4.9\times 10^{7}$\,cm and $n_{H} \sim
1.8-2.6\rm \times 10^{16}\,cm^{-3}$. The former gives
a non-negligible shock height consistently with a
tall shock as suggested from the spin light curve.

\section{Conclusion}

The BeppoSAX and RXTE observations have allowed for the first time the
study of the  broad band X-ray emission spectrum and variability of 
V\,709 Cas which can be characterized as follows:

{\it a)} V\,709 Cas is dominated by the rotational pulsation of the 
accreting WD at a 312.75\,s period, with no sign of orbital or
sideband periodicities. This confirms that it is a disc accretor.

{\it b)} Its spectrum  is hard and well described by
an isothermal optically thin plasma at 27\,keV with complex
absorption
and an iron K\,$_{\alpha}$ fluorescent line, due to reflection
from the WD surface. Evidence of cool and ionized absorbing
material in the pre--shock region is found in this system,  e.g.
from a partial absorber and an iron absorption edge.

{\it c)} The rotational pulsation is compatible with complex 
absorption dominating the low energy range, while at higher energies
pulsation is likely due to the presence of a non--zero shock height
above the accreting poles. 

{\it d)} Variations along the rotational pulse in the partial
covering absorber and reflection are  compatible with
the classical curtain scenario, where accretion material 
flows from the disc towards the polar regions of the WD
via  arc-shaped curtains.

{\it e)} The shape and amplitude of the spin 
light curve as well as the X--ray flux change with time
indicating that V\,709 Cas experiences variations in the
mass accretion rate on timescales  of less than months to years.

\begin{acknowledgements}
We gratefully acknowledge Prof. J. Patterson for the usage of 
the RXTE data. We also wish to thank Dr. C. Done for kindly providing the
reflection fitting code. The BeppoSAX team and Science Data Center staff
are warmly thanked for their help and advice in the data 
handling and reduction. We also acknowledge useful discussion with the
RXTE team, Dr. J. Swank and Dr. K. Jahoda, on PCA instrument
calibration and data analysis. DdM
and GM acknowledge financial support from ASI.

\end{acknowledgements}

%%%%%%%%%%%%%%%%%%%%%%%%%


\begin{thebibliography}{}

 \bibitem []{} Anders E., Grevesse N., 1989, Geochimica et 
Cosmochimica Acta 53, 197

\bibitem {Bea95}
Beardmore A.P., Done C., Osborne J.P., et al.,  1995, MNRAS 272, 749
%Ishida M.,

\bibitem [Beardmore et al. 1998] {Bea98} % Changes in PS of FO Aqr
Beardmore A.P., Mukai K., Norton A.J. et al., 1998, MNRAS 297, 337
%Osborne J.P, Hellier C., 

\bibitem [] {}
Boella  G., Butler R.C., Perola G.C., et al., 1997, A$\&$AS 122, 299

\bibitem [] {}
Bonnet-Bidaud J.-M., Mouchet M., de Martino D., et al., 2001, A\&A,
 in press

\bibitem [] {}
Bradt H.V., Rothschild R.E., Swank J.H., 1993, A\&AS 97, 355

\bibitem [Buckley 1996] {} 
Buckley D.A.H., 1996, In: A. Evans, J.H. Wood (eds.) Proc. 
Cataclysmic Variables and Related Objects, IAU Coll. 158, Space Sci. Lib. 208,
p.185

\bibitem [Buckley 1997] {}
 Buckley D.A.H., Haberl F., Motch C., et al., 1997, MNRAS 287, 117

\bibitem [Deeming75] {deem75}
Deeming T.J., 1975, Ap\&SS 36, 137

\bibitem [Done92] { }
Done C., Mulchaey J.S., Mushotzky R. F., et al., 1992, ApJ 395, 275
 
%Arnaud, K. A.
\bibitem [Done95] { }
Done C., Osborne J.P., Beardmore A.P., 1995, MNRAS 276, 483

\bibitem [de Martino93] {dema93}
 de Martino D., 1993, In: O.Regev, G.Shaviv (eds.) Proc. 2nd Haifa 
Conference on  Cataclysmic Variables and related physics, p.201

\bibitem [demartino95] {dema95}
de Martino D.,  Mouchet M., Bonnet-Bidaud J.M., et al., 1995, A\&A, 298, 849
% Vio R., Rosen S., Mukai K., Augusteijn T., Garlick M., 

\bibitem [demartino99] {dema99}
de Martino D., Silvotti R.,  Buckley D.A.H., et al., 1999, A\&A 350, 517
%G\"ansicke B.T., Mouchet M., Mukai K., Rosen S.R., 

%\bibitem [Dickey90] {DickLock}
%Dickey J.M., Lockman G.J., 1990, ARA\&A 28, 215

\bibitem {} Haberl F., Motch C., 1995, A\&A 297, L37


%\bibitem [Hellier 93] {Hel93}
%Hellier C., 1993, MNRAS 265, L5

\bibitem [Hellier 95] {Hel95}
Hellier C., 1995, In: D.A.H. Buckley, B. Warner (eds.) ASP Conf. Ser. 85,
Cape Workshop on Magnetic Cataclysmic Variables, p. 185


\bibitem [hellier99] {}
Hellier C., 1999, In: C. Hellier, K., Mukai (eds.) ASP Conf. Ser. 157, 
Annapolis Workshop on Magnetic Cataclysmic Variables, p.1



\bibitem {} Inoue H., 1985, SSRv 40, 317

\bibitem [Ishida94] {Ishida}
Ishida M., Mukai K., Osborne J.P., 1994, PASJ 46, L81   

%\bibitem {}
% Kaastra J.S., 1992, An X-Ray Spectral Code for Optically Thin
%Plasmas, Internal SRON-Leiden Report, Version 2.0

\bibitem {}
Kallman T., Mc\,Cray R., 1982, ApJSS 50, 263

\bibitem [] {} Kozhevnikov V.P., 2001, A\&A 366, 891


\bibitem [Matt99] { } 
Matt G., 1999, In: C. Hellier, K. Mukai (eds.) ASP Conf. Ser. 157,
Annapolis Workshop on Magnetic Cataclysmic Variables, p. 299

\bibitem {}
Matt G., Perola G.C., Piro L., 1991, A\&A 247, 25

\bibitem {}
 Mewe R., Kaastra J.S., Liedahl D.A., 1995, Legacy (Journal
of HEASARC) 6, 16

%\bibitem {}
% Mewe R., Lemen J.R., van den Oord G.H.J., 1986, A\&AS 65, 511

\bibitem{} Motch C., Haberl F., Guillout P., et al., 1996, A\&A 307, 459


\bibitem {} Mukai K., In: C. Hellier, K. Mukai (eds.), ASP Conf. Ser. 157,
Annapolis Workshop on Magnetic Cataclysmic Variables, p.33

%\bibitem {} Mukai K., Ishida M., Osborne J.P., 1994, PASJ 48, L87


%\bibitem [Norton et al. 92] {Nort92}
%Norton A.J., Watson M.G., King A.R., Letho H.J., McHardy I.M.
%1992, MNRAS 254 705.

\bibitem [] {} Norton A.J., Beardmore A.P., Taylor P., 1996, MNRAS 280 937

\bibitem [Norton et al. 97] {Nort97}
Norton A.J., Hellier C., Beardmore A.P., et al., 1997, MNRAS 289, 362
%Wheatley P.J., Osborne J.P. \& Taylor P. 


\bibitem [] {} Norton A.J., Beardmore A.P., Allan A., et al., 1999, A\&A
347, 203 

\bibitem [Patterson94] {Patt94}
Patterson J., 1994, PASP 106, 209


%\bibitem [Pattersonetal98] {Pattetal98}
%Patterson J., Kemp J., Richman H.R., et al. 1998, PASP 110, 415


\bibitem [Roberts et al.]{Rob87}
 Roberts D.H., Leh\'{a}r J., Dreher J.W., 1987, AJ 93, 968

%\bibitem {} Rosen S.R., 1992, MNRAS 254, 493

\bibitem [Rosen et al. 1988] {Ros88}
Rosen S.R., Mason K.O., Cordova F.A, 1988, MNRAS 231, 549

\bibitem {}
Singh K. P., White N.E., Drake S.A., 1996, ApJ 456, 766

\bibitem {}
Wynn G., King A., 1992, MNRAS 255, 83

\bibitem {} Warner B., 1986, MNRAS 219, 347

\bibitem [Warner 1995]{war95} % Big CV review
 Warner B., 1995, Cataclysmic Variable Stars, Cambridge
 Univ. Press, Cambridge 

\end{thebibliography}
\end{document}